\documentclass[a4paper,fleqn]{elsarticle}

\usepackage{lineno,hyperref}
\modulolinenumbers[5]
\usepackage{newtxtext}
\usepackage[dvipsnames]{xcolor}
\usepackage{soul}
\usepackage[T1]{fontenc}
\usepackage{ae,aecompl}
\usepackage{graphicx}
\usepackage{amsmath}	
\usepackage{amssymb}
\usepackage{xr}
\usepackage{float}
\usepackage{afterpage} 
\usepackage{floatpag}

\journal{New Astronomy}

\begin{document}

\begin{frontmatter}

\title{Modelling the quenching of star formation activity from the evolution of the colour-magnitude relation in VIPERS\tnoteref{mytitlenote}}
\tnotetext[mytitlenote]{based on observations collected at the European Southern Observatory, Cerro Paranal, Chile, using the Very Large Telescope under programmes 182.A-0886 and partly 070.A-9007.  Also based on observations obtained with MegaPrime/MegaCam, a joint project of CFHT and CEA/DAPNIA, at the Canada-France-Hawaii Telescope (CFHT), which is operated by the National Research Council (NRC) of Canada, the Institut National des Sciences de l'Univers of the Centre National de la Recherche Scientifique (CNRS) of France, and the University of Hawaii. This work is based in part on data products produced at TERAPIX and the Canadian Astronomy Data Centre as part of the Canada-France-Hawaii Telescope Legacy Survey, a collaborative project of NRC and CNRS. The VIPERS web site is \url{http://www.vipers.inaf.it/.}}


\author[ICC,CEA,IDAS,IASF,Bicocca]{G.~Manzoni\corref{correspondingauthor}}
\cortext[correspondingauthor]{E-mail: \texttt{giorgio.manzoni@durham.ac.uk}}

\author[IASF]{M.~Scodeggio}
\author[ICC,IDAS]{C.~M.~Baugh}
\author[ICC,CEA,IDAS]{P.~Norberg}
\author[Trieste]{G.~De~Lucia}
\author[Fritz]{A.~Fritz}
\author[Chile,Brera]{C.P.~Haines}
\author[Bologna]{G.~Zamorani}
\author[IASF]{A.~Gargiulo}
\author[Statale,Brera,NewGigi]{L.~Guzzo}
\author[Brera]{A.~Iovino}
\author[Poland]{K.~Ma{\l}ek}
\author[Poland,Pollo]{A.~Pollo}
\author[Barcelona,Poland]{M.~Siudek}
\author[Bologna]{D.~Vergani}

\address[ICC]{Institute for Computational Cosmology (ICC), Department of Physics, Durham University, South Road, Durham DH1 3LE, UK}

\address[CEA]{Centre for Extragalactic Astronomy (CEA), Department of Physics, Durham University, South Road, Durham DH1 3LE, UK}
\address[IDAS]{Institute for Data Science (IDAS), Durham University, South Road, Durham DH1 3LE, UK}
\address[IASF]{INAF - Istituto di Astrofisica Spaziale e Fisica Cosmica Milano, via Bassini 15, 20133 Milano, Italy}
\address[Bicocca]{Dipartimento di Fisica G. Occhialini, Universit\`a di Milano-Bicocca, Piazza della Scienza 3, I-20126 Milano, Italy}
\address[Trieste]{INAF - Osservatorio Astronomico di Trieste, via G. B. Tiepolo 11, 34143 Trieste, Italy}
\address[Fritz]{OmegaLambdaTec GmbH, Lichtenbergstra{\ss}e 8, 85748 Garching, Deutschland}
\address[Chile]{Instituto de Astronom{\'i}a y Ciencias Planetarias de Atacama, Universidad de Atacama, Copayapu 485, Copiap{\'o}, Chile}
\address[Brera]{INAF - Osservatorio Astronomico di Brera, Via Brera 28, 20122 Milano --  via E. Bianchi 46, 23807 Merate, Italy}
\address[Bologna]{INAF - Istituto di Astrofisica Spaziale e Fisica Cosmica Bologna, via Gobetti 101, I-40129 Bologna, Italy}
\address[Statale]{Dipartimento di Fisica ``Aldo Pontremoli'', Universit\`{a} degli Studi di Milano, via G. Celoria 16, 20133 Milano, Italy}
\address[NewGigi]{INFN - Sezione di Milano, via G. Celoria 16, 20133 Milano, Italy}
\address[Poland]{National Centre for Nuclear Research, ul. Hoza 69, 00-681 Warszawa, Poland}
\address[Pollo]{Astronomical Observatory of the Jagiellonian University, Orla 171, 30-001 Cracow, Poland}
\address[Barcelona]{Institut de F\'{\i}sica d'Altes Energies (IFAE), The Barcelona Institute of Science and Technology, 08193 Bellaterra (Barcelona), Spain}

\begin{abstract}
We study the evolution of the colour-magnitude relation for galaxies in the VIMOS Public Extragalactic Redshift Survey (VIPERS) by introducing the concept of the bright edge, and use this to derive constraints on the quenching of star formation activity in galaxies over the redshift range $0.5 < z < 1.1$.
The bright-edge of the colour-magnitude diagram evolves with little dependence on galaxy colour, and therefore on the amount of star formation taking place in bright galaxies. We modelled this evolution with delayed exponential star formation histories (SFHs), to better understand the average time-scale of the turn-off in star formation activity. We show that using SFHs without quenching, the transition from the blue cloud to the red sequence is too slow. This indicates that a scenario purely driven by the consumption of the gas inside each galaxy does not reproduce the observed evolution of the colour-magnitude bright edge. For instantaneous quenching, the best match to the observations assumes that galaxies stop their star formation at a randomly distributed time up to $\sim 2 - 2.5$ Gyr after observation. We argue that quenching is required over a wide range of stellar masses. 
Qualitatively similar evolution of the bright edge is found in the predictions of a semi-analytical galaxy formation model, but quantitatively there are marked differences with the observations. This illustrates the utility of the bright edge as a test of galaxy formation models. The evolution changes and no longer matches the observed trend if feedback from heating by active galactic nuclei is turned off.  
\end{abstract}

\begin{keyword}
Cosmology: observations \sep Cosmology: large scale structure of Universe \sep Galaxies: distances and redshifts \sep Galaxies: statistics
\end{keyword}

\end{frontmatter}
\section{Introduction}
\label{sec:intro}
Over the past two decades galaxy evolution studies have provided us with fundamental insights into galaxy formation. The global star formation rate density in the Universe peaked at a redshift of $z \sim 2$, and has then steadily declined, by an order of magnitude, to the present day  (\cite{madau96,lilly96}; see \cite{madau_dickinson_14} for a comprehensive review). This decline is associated with the gradual transfer of star formation activity from more massive galaxies at high redshift to progressively less massive ones over cosmic time, an effect referred to as downsizing \cite{cowie96,gavazzi_scodeggio_96,thomas05,treu05,juneau05,siudek17}. This means that by today, smaller and smaller galaxies have experienced a star-forming phase and then moved into a passive stage. This results in a progressive extension of the passive galaxy population towards lower stellar masses \cite[see for example][]{delucia07,kodama07,rudnick09}, as star-forming ``blue cloud'' galaxies migrate to the quiescent, passively evolving ``red sequence''. 
The observed bimodality in many photometric, spectroscopic, and morphological galaxy properties \cite[see, for example, ][]{strateva01,baldry04,balogh04,kauffmann03,krywult17} has been put forwards as an indication that the transition between the red and blue populations might take place quite rapidly \cite[see for example][]{faber07}. This would imply the existence of a physical process capable of suppressing the star formation activity that operates on a much shorter time-scale than that on which the gas is consumed inside a galaxy. We refer to this physical process as {\it quenching}. 

The origin of the star formation  quenching remains controversial and may not be due to one process. One of the first mechanisms proposed to quench star formation in galaxies was ram pressure stripping \cite{gunn_gott_72}, but its effectiveness appears to be limited to clusters of galaxies,  based on where we have been able to identify galaxies suffering ram pressure stripping \cite[see for example][]{giovanelli_haynes_85}. This ``galaxy strangulation'' quenching mechanism has received renewed attention as the possible primary driver for star formation quenching \cite{peng15}. With the advent of galaxy formation simulations, a simple AGN feedback model was introduced to shut down gas cooling in massive halos, in order to match the bright end of the observed galaxy luminosity function  \cite{benson03,croton06,bower06,delucia07}. Observationally the effectiveness of AGN feedback in shutting off star-formation has been studied in \cite{Vergani18} using the NUV$rK$ diagram for VIPERS galaxies with stellar masses greater than $5\times10^{10} M_\odot$. 
However, the observational evidence for such feedback, in particular for the range of halo masses over which it is required to be effective 
is still unclear  \cite{bongiorno16,taylor_kobayashi_16}. \cite{Henriques:2017} argued that the star formation quenching in their semi-analytical model produces predictions that agree qualitatively with observations, but is too effective in dense regions and predicts too much recent star formation in massive galaxies. A similar conclusion was reached by \cite{Bluck:2016} on comparing galaxy formation models with Sloan survey observations.

From the observational point of view, even establishing that quenching is taking place inside previously star-forming galaxies is not trivial. 

E+A galaxies \footnote{The spectra of $\rm{E+A}$ galaxies have strong Balmer absoprtion features but do not show evidence of ongoing star formation, such as [OII] or H$\alpha$ emission lines. This suggests that they experienced a starburst about $1$Gyr prior to observation, and so do not have the canonical spectrum of an elliptical galaxy, which are typically assumed to have had no star formation over a much longer period, but rather that of an elliptical combined with an A-star. They are also referred to as post-starburst galaxies; see \cite{dressler_gunn_83,wild09}.} experienced quenching in their recent past, but their rarity, while lending support to the hypothesis of a short time-scale transition from the blue cloud to the red sequence, makes the understanding of both their connection with the wider galaxy population, and of the details of the quenching process, complicated \cite[see for example][]{kaviraj07,yesuf14,wild16}. 
Similar uncertainties affect the study of the ``green-valley'' galaxies \cite{wyder07,martin07}, which have been considered either as a ``normal'' evolutionary stage common to all galaxies \cite{martin07,salim14}, or as ``peculiar'' objects representative of the evolution of a small fraction of the overall galaxy population \cite{smethurst15}. \cite{schawinski14} claim instead that the green valley is just populated by normal star-forming galaxies at very high masses. A number of studies have attempted to model the properties of the overall galaxy population to derive constraints on the quenching of star formation \cite[see for example][]{CIESLA_16,abramson16,lian16,Bluck:2016,Davies:2019}. The main results are a confirmation of a short time-scale for the quenching (on the order of 200 to 500 Myr, \cite{CIESLA_16}), and an estimate that quenching affects a relatively large fraction of galaxies, around 30 to 45$\%$ of the overall population \cite{lian16}. However, \cite{abramson16} argue that the very idea of quenching comes from the use of over-simplified canonical parametrizations of the star formation history (SFH) which in general are not able to reproduce both the tail of high star formation rate (SFR) at $z\sim1$ and the low SFRs seen today.

Here, we take 65\,000 galaxies from the VIPERS galaxy redshift survey \cite{guzzo14} to study the evolution of the colour-magnitude relation for the overall galaxy population. In particular, we use the evolution of the ``bright edge" of the galaxy distribution in the rest-frame $U-V$ colour vs. absolute $V$-band magnitude plane to derive illustrative constraints on the SFH quenching time-scale, and on the ability of this process to affect galaxies over a large range of stellar masses. The large number of galaxies in VIPERS allows us to develop a statistical understanding of quenching time-scales. In terms of a single galaxy, the quenching is implemented as the instantaneous truncation of star formation activity. Given the depth of VIPERS our study primarily focuses on bright galaxies, so we are unable to draw conclusions about quenching in faint galaxies (see e.g. \cite{Davies:2019} for constraints on quenching in faint galaxies at somewhat lower redshifts than those considered here). 

This paper is set out as follows: in Section~\ref{sec:data} we describe the modelling and assumptions made to obtain galaxy properties from the VIPERS data; in Section~\ref{sec:cm_evolution} we explain how we model different SFH scenarios; in Section~\ref{sec:cm_relation} we present the observed colour-magnitude relation and its evolution tracked using the bright edge concept; in Section~\ref{sec:results} we show the results of our synthetic evolution compared to the observed sample to constrain quenching; in Section~\ref{sec:galform} we compare our findings with the evolution of the colour-magnitude relation predicted by the {\tt GALFORM}  semi-analytic models and, finally, in Section~\ref{sec:discussion} we discuss our main results.

We use a flat $\Lambda$ CDM cosmology with $\Omega_{\rm m} = 0.3$, $\Omega_{\Lambda}=0.7$ and $H_0 = 70 \; \rm{km}/\rm{s}/\rm{Mpc}$, unless stated otherwise.

 \begin{figure*}
   \centering
    \includegraphics[width=\textwidth]{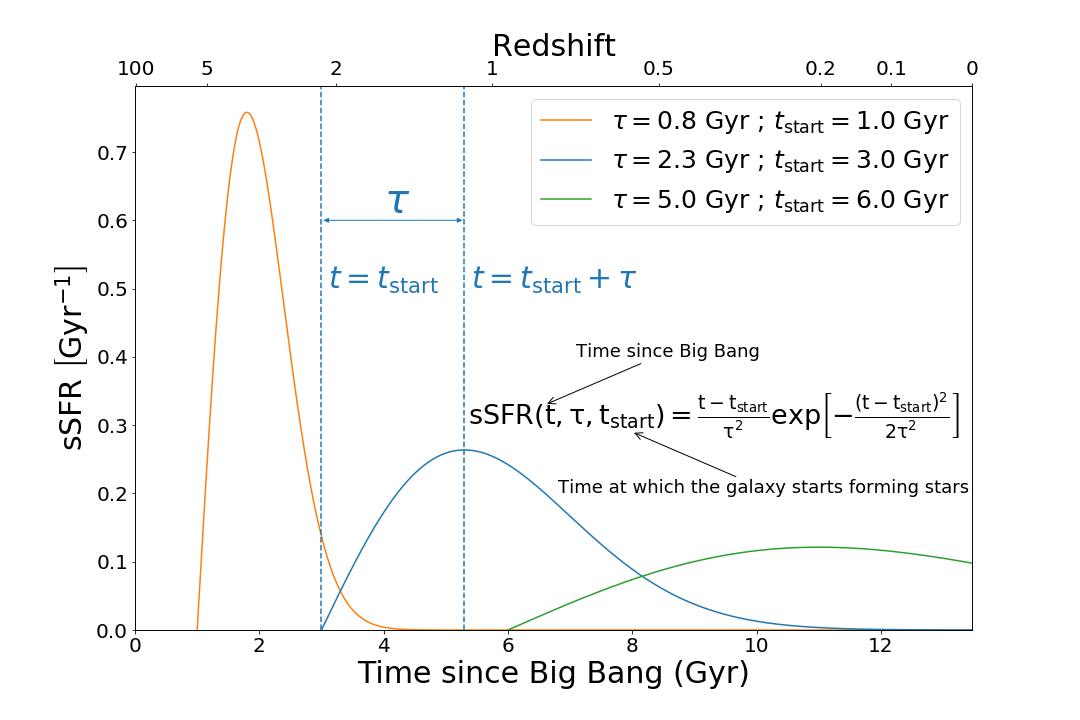}
     \caption{Examples of delayed exponential SFHs for different $\tau$, offset by $t_{\rm{start}}$ from the Big Bang, which corresponds to $t=0$. Through SED fitting, which is based on photometric and spectroscopic data (see \cite{FRANZETTI_005}), every galaxy is assigned an age, $t-t_{\rm{start}}$, and a characteristic time-scale of star-formation, $\tau$, which determine the form of the SFH. We stress that each SFH plotted has a different $t_{\rm{start}}$, equivalent to the redshift at which the galaxy starts to form stars.}
\label{fig:SFHs}
   \end{figure*}

\section{Modelling of galaxy properties from the data}
\label{sec:data}

\subsection{The VIPERS data}

The galaxy sample used here is based on the full data release of the VIPERS spectroscopic survey. VIPERS is a galaxy redshift survey carried out with the VIMOS spectrograph at the ESO Very Large Telescope (see \cite{guzzo14} for a full description of the survey), covering approximately $23.5$ deg$^2$ of sky, and targeting a sample of galaxies brighter than $i_{\rm AB}=22.5$, selected to be at redshift $z>0.5$ on the basis of a simple, but rather effective, colour-colour selection criterion. 

The full VIPERS data release \cite{scodeggio18} provides a spectroscopic catalogue with redshift measurements for $86,775$ galaxies, coupled with a parent photometric catalogue largely based on the VIPERS Multi-Lambda survey \cite{moutard16a}, providing photometric data that include the GALEX FUV and NUV bands, the CFHTLS data release T0007 $u, g, r, i$ and $z$ bands, and the WIRCAM or VIDEO infrared $K_{\rm s}$ band (see \cite{scodeggio18} and \cite{moutard16a} for details).

Here we use the subset of galaxies with a reliable redshift measurement (i.e. a measurement with a probability larger than 95\% of being correct, corresponding to a reliability flag $\geq 2$\ in the VIPERS catalogue, see \cite{scodeggio18} for details), limited to the redshift range $0.5 < z < 1.1$, giving a total of $64,889$ objects. This subset is fully representative of the galaxy population over this redshift range, thanks to two important properties of the VIPERS survey: the high sampling rate achieved by VIPERS, whereby on average 47\% of the complete parent photometric sample has been observed, and the very high success rate for the redshift measurements, with on average 90\% of the targeted objects having a redshift measurement (note that the spectroscopic success rate is only weakly dependent on galaxy properties, as shown by Fig.~7 of \cite{scodeggio18}).

\subsection{Galaxy properties and SED fitting}
\label{sec:absmags_and_masses}

Here, we are interested in studying the observed evolution of the bright edge of the $V$-band rest frame absolute magnitude vs $U-V$ rest frame colour. These quantities come from the Public Data Release 2 (PDR-2) which is available at \url{http://vipers.inaf.it/rel-pdr2.html} and presented in \cite{scodeggio18}. To analyse the redshift evolution of the colour-magnitude relation, we divide the sample into six redshift bins, with a width of 0.1, from $z=0.5$ to $z=1.1$ (which corresponds to a look-back time which ranges between approximately $5$ and $8$ Gyr).

We reconstruct the SFH of individual galaxies using SED fitting carried out with the {\tt GOSSIP} software \cite{GOSSIP_26}. We start from the $ugrizKs$ photometry, supplemented with the spectroscopic data, in order to reduce the well known degeneracies between age, star formation timescale and dust extinction that afflict SED fitting results derived  purely from photometric data \cite[see the discussion in][]{rthomas17}. We use a template library based on the {\tt PEGASE 2} model  \cite{PEGASE_10}, assuming a delayed exponential SFH according to the prescription from \cite{GAVAZZI_25}\footnote{We write the SFH in terms of the specific SFR (sSFR $ = \rm{SFR} / M$) as in our model the SFH is normalised to produce $1 M_{\odot}$ of stars.}:
\begin{equation}
    \label{eq:sfh_sandage}
    \rm{sSFR}(t,\tau,t_{\rm{start}}) = \frac{t-t_{\rm{start}}}{\tau^2}\exp \left[ -\frac{(t-t_{\rm{start}})^2}{2\tau^2}\right],
\end{equation}
where $t$ is the cosmic time (with $t=0$ corresponding to the Big Bang), $t_{\rm{start}}$ is the moment at which the galaxy starts to form stars and $\tau$ is the characteristic timescale of star formation which identifies the maximum of the sSFR. This form is often referred to as a SFH \textit{``a la Sandage''} as \cite{SANDAGE_04} was the first to discuss  such a SFH. Fig.~\ref{fig:SFHs} shows some example SFHs generated using this parametrization to illustrate the influence of the parameters on the SFH.  

The SFH parameters that we constrain in our SED fitting are the characteristic time-scale, $\tau$, and the age of the galaxy, $t-t_{\rm{start}}$ (since $t$ is related to redshift, $t_{\rm{start}}$ can also be deduced). The benefit of using a delayed exponential SFH, as in Eq.~\ref{eq:sfh_sandage}, is that such a SFH displays an initial increase of the SFR that peaks at $t-t_{\rm{start}}=\tau$ (see Fig.~\ref{fig:SFHs}), so that late-type galaxies which are still actively forming stars can be described as well as passive early-type galaxies: late-type galaxies will tend to be fitted with larger values of $\tau$ than early-types. Recall that a small value of $\tau$ means that the majority of stars are formed in the early stages of the life of the galaxy, which is usually the case for early-types. 

We have chosen to use the {\tt PEGASE} model because it can compute galaxy SEDs with self-consistent evolution of the metallicity and the internal extinction, driven by the input SFH. The template library used here covers a grid of galaxy ages, $t-t_{\rm{start}}$, from 0.1 to 15 Gyr with a step of 0.1 Gyr, and a grid of star formation time-scales, $\tau$, ranging from 0.1 to 25 Gyr, again with a step of 0.1 Gyr. For each galaxy in our sample we estimate an age, $t-t_{\rm{start}}$, and a star formation time-scale, $\tau$, based on the value of these two quantities for the best-fitting template, with the constraint that the galaxy age is less than the age of the Universe at the redshift of the galaxy. We have checked the SFRs inferred from our best-fitting SFHs are in general agreement with observationally inferred SFRs (\ref{sec:sfh}).

The delayed exponential SFH model is undoubtedly an over simplification of the actual SFH history in galaxies. Several studies have compared simple empirical SFH models with the output of physical models of galaxy formation \cite{Pforr:2012, gladders13, mitchell13, simha14, diemer17}. \cite{mitchell13} showed that simple, declining exponential SFHs, when used to fit the photometry of {\tt GALFORM} galaxies, could nevertheless give a reasonable estimate of properties such as stellar mass. \cite{simha14} compared a wider range of parametric SFHs to those predicted in a hydrodynamic simulation of galaxy formation. They found that a simple exponential SFH gave systematic errors in galaxy colours. Their ``lin-exp'' model, which is equivalent to the delayed-exponential used here, performed much better overall, experiencing problems mainly for the very bluest and reddest galaxies.

\section{Modelling the evolution of the colour-magnitude relation}
\label{sec:cm_evolution}

The modelling of SFHs discussed in Section~\ref{sec:data} provides us with a tool to predict quantitatively the evolution of the colour-magnitude relation as a function of redshift for the galaxies in our sample. The quality of this modelling is tested in \ref{sec:sfh}, in which we  compare the SFR predicted by our model with that inferred from the luminosity of the [OII]$\lambda$3727 emission line. We follow the recipe of \cite{MOUSTAKAS_14} to estimate the SFR from the luminosity of the [OII]$\lambda$3727 emission line, using the luminosity of the line and the rest-frame B-band absolute magnitude, as given in Eq.~\ref{eq:SFR_OII}. This estimate is completely independent from our SED fitting modelling. In \ref{sec:sfh} we show that the SFR vs $U-V$ colour and the SFR vs redshift relations, whilst yielding different absolute values, follow the same trend for the two estimates. This gives us some confidence that the SED modelling is giving a reliable prediction of the SFR at the epoch of the observation.

In this section we describe the different scenarios (Fig.~\ref{fig:Sketch_SFH_99_0}) that we test to predict the photometric evolution of galaxies found in a given redshift bin.  
We then compare this predicted evolution with the observed distribution within different redshift bins, to select which of our SFH scenarios best describes the observed galaxies, and thereby derive constraints on the quenching of star formation activity.

\subsection{Model description}

We use a combination of observations and predictions based on our SED fitting results to estimate the amount of photometric evolution the galaxies in our sample are most likely to undergo. The starting point is the set of observed properties for galaxies in a given redshift bin (which we call the start-redshift bin), while the end point is the predicted properties for these objects at some later time (the end-redshift epoch). First, we associate with each galaxy in the start-redshift sample an evolutionary model expressed in terms of a SFH, by selecting the model which provides the best-fitting SED template to the galaxy's observed broad band magnitudes and spectrum. In doing so we assume that the evolutionary model provides a good description of the galaxy properties for an extended time around the cosmological epoch when the galaxy was observed. Since this model provides us with a synthetic SED at {\it all} possible times (from the start of star formation activity, assumed to take place at z=4, to the present epoch), we can use it to predict how much the galaxy luminosity changes with time in the various photometric bands. In particular, we compute the change in luminosity (for each band) after a time interval corresponding to the difference in look-back time between end and start-redshifts ($\Delta L_{\rm model}$), and we use these predictions to compute the expected absolute magnitudes and rest-frame colour of the galaxy at the end-redshift epoch ($L_{\rm zend} = L_{\rm obs} + \Delta L_{\rm model}$, with $L_{\rm obs}$ being the luminosity at the epoch of the observation, and $L_{\rm zend}$ the predicted luminosity at the end-redshift epoch).\footnote{We caution the reader that we have used different SED templates from those used to derive properties in the VIPERS database, and so the absolute values of properties that we obtain should not be compared to those in the database. 
However, since we are primarily interested in the variation of fitted model parameters with redshift rather than their absolute values, this is not an issue for our analysis. }


\begin{figure}
   \centering
    \includegraphics[width=0.74\textwidth]{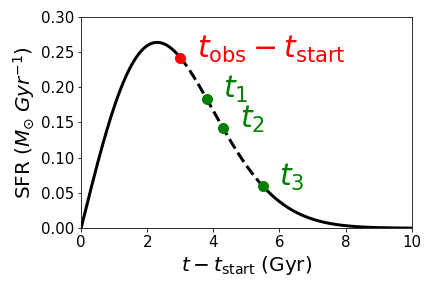}
    \includegraphics[width=0.74\textwidth]{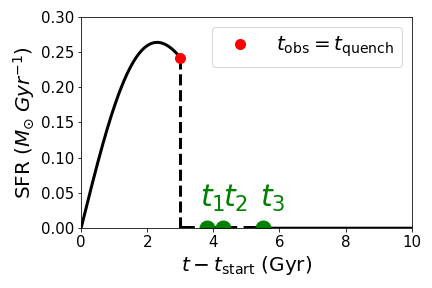}
    \includegraphics[width=0.74\textwidth]{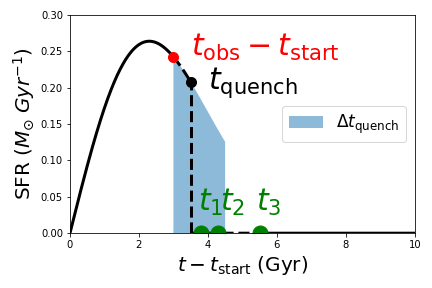}
     \caption{Schematic representation of the three SFH scenarios considered. \textit{Top panel}: smooth delayed exponential SFH (as in Eq.~\ref{eq:sfh_sandage}), without any truncation or quenching. \textit{Middle panel}: SFH that is quenched at the redshift at which the galaxy is observed i.e. the subsequent SFR is set to zero. \textit{Bottom panel}: SFH that is  quenched at a randomly selected time within a time interval (shown by the shaded area) after the epoch of observation. The x-axis indicates the age of the galaxy. $t_1$, $t_2$ and $t_3$ represent the time at which we test the synthetic evolution of the colour magnitude relation (see Section~\ref{sec:results}). We consider different values for the time interval, $\Delta t _{\rm quench}$ over which quenching can take place.}
         \label{fig:Sketch_SFH_99_0}
   \end{figure}

To test other SFH scenarios we add the possibility that the SFH is truncated or quenched. The quenching is always assumed to be both complete (i.e. the SFR is set equal to zero after the quenching), and instantaneous (limited to the time-resolution of the models). These two assumptions are extreme but allow us to treat the problem in a simple way and to retrieve general trends rather than detailed conclusions. We have experimented with two possible quenching scenarios: the first in which all galaxies are quenched at the same time, immediately after the start-redshift epoch (see the middle panel of Fig.~\ref{fig:Sketch_SFH_99_0} for an example of such a SFH), and the second where the quenching time is drawn uniformly from a limited time interval (bottom panel of Fig.~\ref{fig:Sketch_SFH_99_0}). We have explored three possibilities, specified by the length of the time interval from which the quenching time is selected: $1$, $1.5$, and $2.5$ Gyr. In all cases, the start of the period over which quenching could occur starts immediately after the start-redshift epoch. The reason for the choice of these three interval times is that they are representative of the range of lookback times of the VIPERS survey redshift bins. By varying the time range over which quenching can take place, we vary the fraction of galaxies that are quenched at a given redshift. Of course, for end-redshifts corresponding to a change in time interval that is greater than the time interval from which the quenching epoch is selected, all galaxies will be quenched. 

\subsection{Example tracks for the no-quenching case}

\begin{figure}
   \centering
   \includegraphics[width=0.8\textwidth]{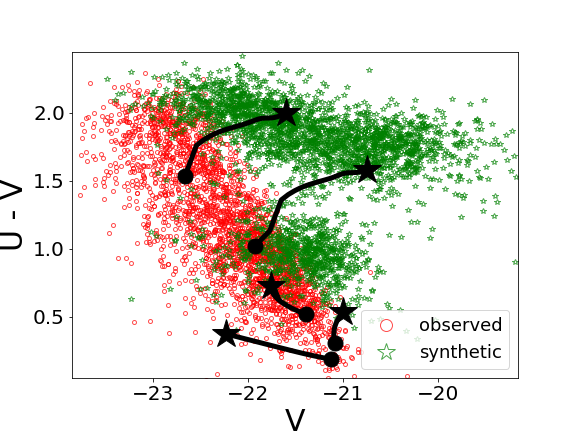}
   \caption{Evolution of the colour-magnitude relation starting from the (observed) start-redshift bin $1.0<z<1.1$ (red empty circles) over a period of $2.5$ Gyr, terminating  in the end-redshift bin $0.5<z<0.6$ (green empty stars). The evolution is based on the best-fitting {\tt PEGASE} models built with a delayed exponential SFH (see the  schematic representation in the top panel of Fig.~\ref{fig:Sketch_SFH_99_0}, with the green points being plotted at $t_3$, which corresponds to the end-redshift). 
   Thick black lines show the evolutionary tracks of a few ``example galaxies'' which start at the filled circle and end at the filled star.
   }
   \label{fig:EVOLUTION}
\end{figure}

Fig.~\ref{fig:EVOLUTION} shows an example of the evolution expected in the colour-magnitude plane when galaxies follow the basic "no-quenching SFH" illustrated in the top panel of Fig.~\ref{fig:Sketch_SFH_99_0}. 
In Fig.~\ref{fig:EVOLUTION} we highlight, using thick solid lines, the evolution of SFH tracks in the colour-magnitude plane for a few example galaxies over the whole time interval of $2.5$ Gyr (corresponding to the evolution from the median redshift in the $1.0<z<1.1$ start-redshift bin to the median redshift in the $0.5<z<0.6$ end-redshift bin). 
The points in the background of Fig.~\ref{fig:EVOLUTION} are the observed galaxies in the start-redshift bin (red circles) and their evolved counterparts in the end-redshift bin (green stars).

The reddest of the example galaxies ($U-V \sim 1.5$, $V \sim -22.6$)  has already completed its star formation life cycle, and undergoes purely passive evolution, fading in $V$ magnitude and becoming redder still in $U-V$. The progressively bluer galaxies can be  characterised by increasingly more important star formation activity, with the three bluest objects ($U-V < 0.6$) showing how a combination of different ages and star formation time-scales can result in significantly different evolutionary tracks for galaxies starting with similar properties at the epoch of their observation. If we now consider the population of galaxies instead of individual objects, the overall evolution that we see is that of a global move towards redder colours and fainter magnitudes, creating a more defined bimodality between the blue cloud and the red sequence (see the deficit of green stars around $U-V \sim 1.5$): galaxies in the redder half of the rest-frame colour distribution at the start-redshift epoch evolve to reach the red sequence at the end-redshift epoch, while galaxies that start in the bluest part of the distribution can remain in part of the blue cloud, albeit with significantly redder colours. We stress that this behaviour is not obvious when examining only a few example tracks but is something that becomes apparent when considering the population of galaxies. For example, the bluest object in the example tracks becomes brighter as it is evolved from its observed colour and magnitude. Other galaxies near the bright edge, however,  are predicted to evolve such that they become fainter and redder. These objects drive the bimodalilty in the evolved colour - magnitude relation. Note that we do not require that the evolved population meets the VIPERS selection, so some evolved galaxies would not be observed in VIPERS.

\section{The evolution of the colour-magnitude relation in VIPERS}
\label{sec:cm_relation}

To test our modelling of  galaxy SFHs using the colour -- magnitude relation, we need to choose a robust feature we can use to quantitatively track and describe the evolution of the VIPERS galaxies. We decide to define a bright edge in the colour magnitude distribution. 
Due to the declining  cosmic star formation rate density with increasing cosmic time, we expect a general reduction in the luminosity of the brightest galaxies which can be tracked by the bright edge. 
Of course, both luminosity and colour can change in response to different SFHs, as shown in Fig.~\ref{fig:EVOLUTION}. However, since studies like \cite{davidzon13} have already focused on galaxy colours, we have decided instead to devote our attention to studying the evolution in galaxy magnitudes within the colour-magnitude plane.

\subsection{Defining the bright edge of the colour -- magnitude relation}
\label{sec:edge_measurement}

   \begin{figure}
   \centering
    \includegraphics[width=0.8\textwidth]{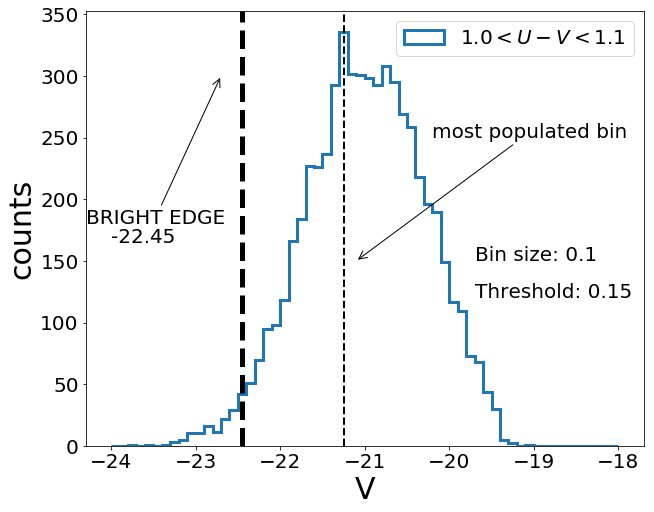}
     \caption{Histogram of $V$-band absolute magnitudes in the colour bin $1.0<U-V<1.1$ for the VIPERS sample, along with the details of the operational definition of the bright edge. The thin dashed vertical line indicates the most populated magnitude bin and the thick dashed vertical line identifies the location of the bright edge in that colour bin, where the differential galaxy counts drop below $15\%$ of that in the most populated bin. }
         \label{fig:histo_bright_edge}
   \end{figure}

To define the bright edge of the colour -- magnitude relation in an objective and quantitative way, we partition the plane into a grid of cells 0.10 mag wide along both axes. For a fixed colour bin we count the number of galaxies in each absolute magnitude bin and find the most populated bin. Moving from this bin in the direction of brighter magnitudes, we define the edge location as the first bin for which the occupancy falls below $15\%$ of that in the most populated bin (see Fig.~\ref{fig:histo_bright_edge} for an illustration of this process for one colour bin). This process is repeated for each colour bin. 

Since the values of the bin size and occupancy threshold are arbitrary, we have checked that our results are not significantly affected by these choices. Specifically, we have tested the robustness of the edge definition against the size of the bins in both the rest-frame colour and absolute magnitude, and against the occupancy threshold with respect to the most populated bin. The typical variation of the bright edge location is approximately 0.15 mag.  
This value receives contributions from two sources: 1) the statistical uncertainty arising from the number of galaxies in the bins used in the colour magnitude plane (we varied the rest-frame colour bin size between 0.06 and 0.22 mag, and the absolute magnitude bin between 0.04 and 0.20 mag) and 2) a systematic error due to the choice of occupancy threshold (which we varied between 1\% and 50\%).
We therefore consider 0.15 mag as the uncertainty in the location of the bright edge, to be compared with the observed evolution of approximately 1.0 mag (for a U-V colour of 1.25, at the centre of the colour distribution) across the redshift range covered by our data. The relative insensitivity to the choice of the occupancy threshold also ensures that the edge definition is insensitive to the details of the luminosity distribution within the different colour bins (like, for example, the faint-end slope of the luminosity function, which, in turn, depends on galaxy colour). 

In \ref{app:galform} we carry out a test to check if the number density of galaxies has any effect on the definition of the edge. Specifically, in Fig.~\ref{fig:galform_zbins} we draw the bright edge in the colour-magnitude which comes from the \cite{gonzalez14} model which makes use of the GALFORM semi-analytic code. Sub-sampling the data to the number of VIPERS galaxies in every analogue redshift bin (top panel of Fig.~\ref{fig:galform_zbins}) does not affect the location of the bright edge in a systematic way.

\subsection{The evolution of the edge of the colour -- magnitude relation}
\label{sec:edge_evolution}
 
  \begin{figure*}
   \centering
    \includegraphics[width=\textwidth]{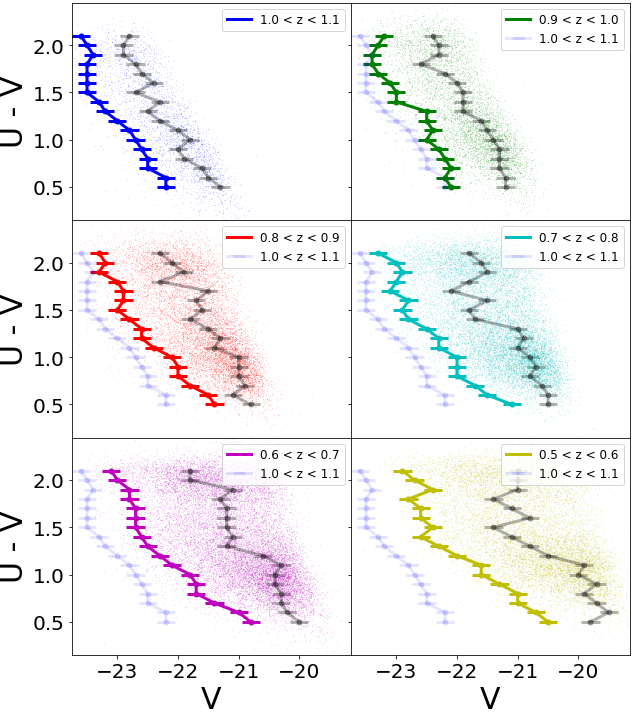}
     \caption{The colour-magnitude relation for the VIPERS sample, split into six redshift bins as labelled on each panel. The thick solid lines mark the bright edge of the distribution in the various bins, with an indicative  uncertainty of 0.15 mags shown by the horizontal error bars} (see Section~\ref{sec:edge_measurement} for the operational definition of the location of the bright edge). The grey thick lines mark the most populated magnitude bin for each colour bin. The faint blue line is the bright-edge from the highest redshift bin that is reproduced in every panel for comparison. 
\label{fig:cm_edge_evolution}
   \end{figure*}

\begin{figure*}
   \centering
   \includegraphics[width=\textwidth]{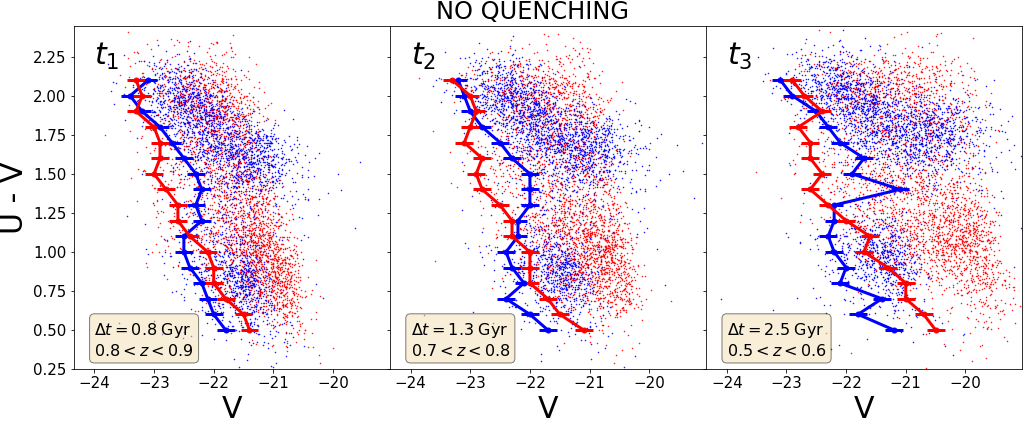}
\caption{Colour -- magnitude relations for the synthetically evolved population (blue points) in three arrival or end redshifts (corresponding to $t_1$, $t_2$ and $t_3$ in Fig.~\ref{fig:Sketch_SFH_99_0}) and of the observed VIPERS data in each end redshift bin (red points). The bright edge location of each sample is drawn using the same bins in colour. The evolution is based on the smooth delayed exponential SFH which characterises the no-quenching scenario illustrated in the top panel of Fig.~\ref{fig:Sketch_SFH_99_0}. The horizontal error bars show an indicative average error of 0.15 mag}.
   \label{fig:EVOLUTION_noquench}
\end{figure*}

\begin{figure*}
\centering
\includegraphics[width=\textwidth]{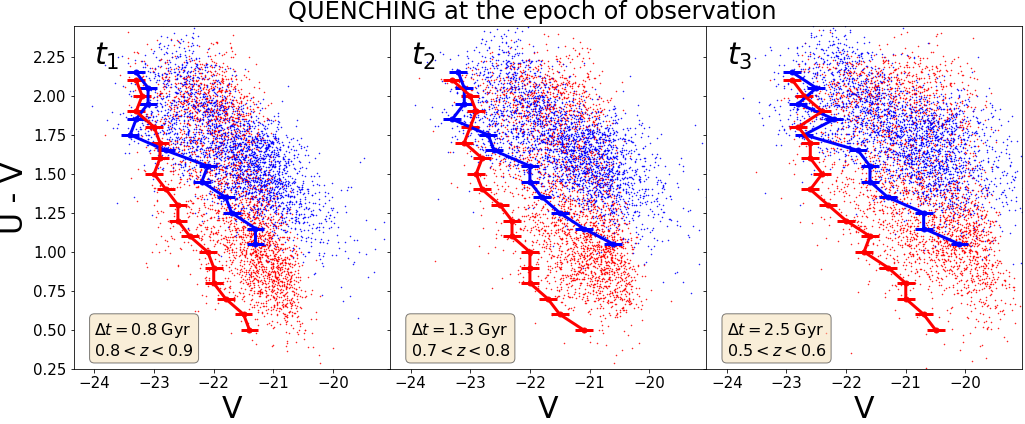}
\caption{Same as Fig.~\ref{fig:EVOLUTION_noquench} but with evolution computed using an instantaneously quenched delayed exponential SFH (i.e. SFR set to zero at the redshift at which the galaxy is observed). This SFH scenario corresponds to the schematic representation in the middle panel of Fig.~\ref{fig:Sketch_SFH_99_0}. The horizontal error bars show an indicative error of 0.15 mag}. 
\label{fig:EVOLUTION_quench0}
\end{figure*}
\begin{figure*}
\centering
\includegraphics[width=\textwidth]{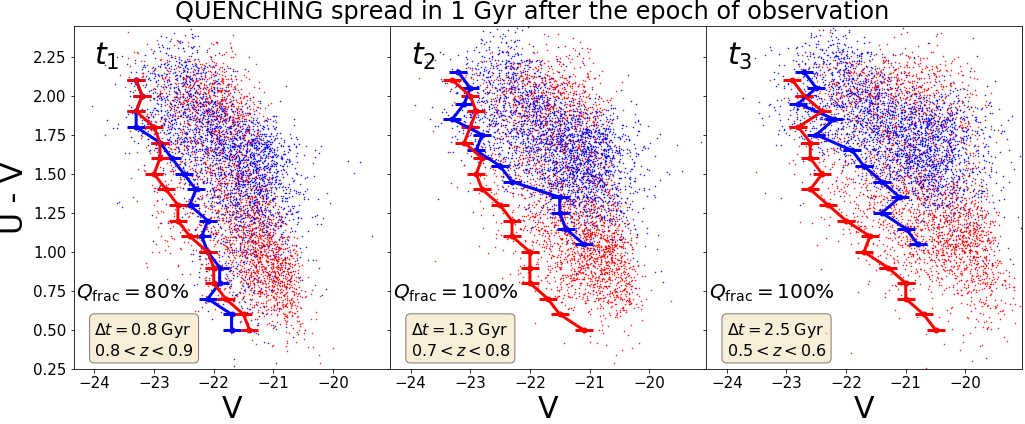}
\includegraphics[width=\textwidth]{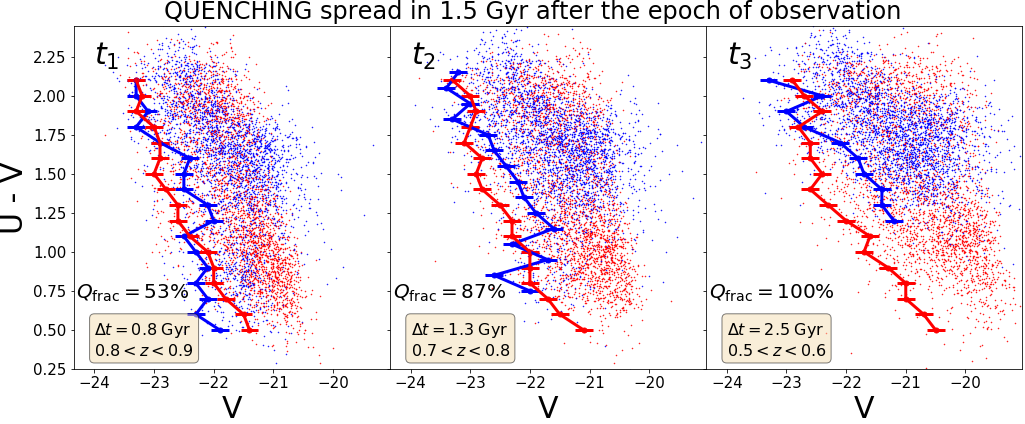}
\includegraphics[width=\textwidth]{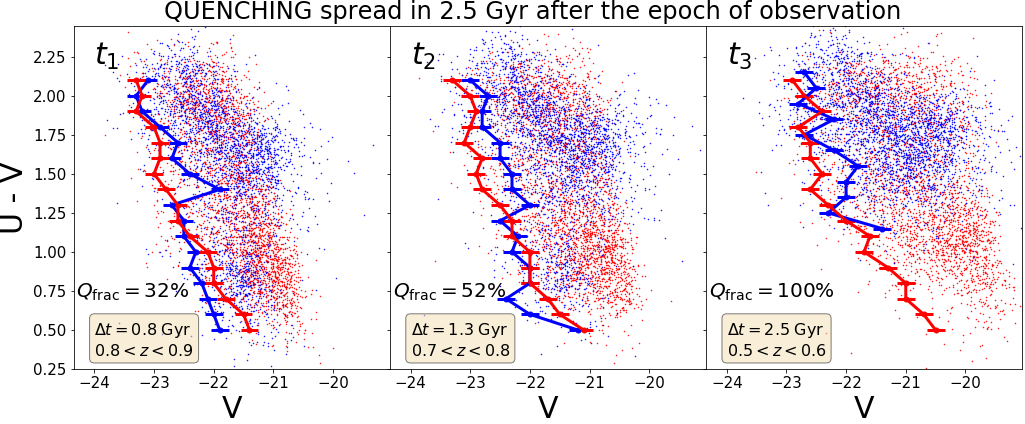}
\caption{Same as Fig.~\ref{fig:EVOLUTION_noquench} but considering different quenching scenarios for the SFH. This time the evolution is based on the ``delayed-quenching'' scenario which means that for each galaxy the SFR is set to zero at a random time within a time interval after the epoch of observation of duration $\Delta t_{\rm{quench}}$. Three different $\Delta t_{\rm{quench}}$ values are explored: 1 Gyr (top panel), 1.5 Gyr (middle panel) and 2.5 Gyr (bottom panel). A schematic representation of the delayed-quenched SFH is shown in the bottom panel of Fig.~\ref{fig:Sketch_SFH_99_0}, with the shaded area representing $\Delta t_{\rm{quench}}$ of 1.5 Gyr. The fraction of galaxies that is quenched, $Q_{\rm{frac}}$, at each end-redshift is written in each panel (see Section~\ref{sec:quenching_scenarios} for a  discussion of how $Q_{\rm{frac}}$ is calculated and interpreted). The horizontal error bars show an indicative error of 0.15 mag}.
\label{fig:EVOLUTION_quench123}
\end{figure*}

In Fig.~\ref{fig:cm_edge_evolution} the thick solid lines mark the bright edge of the galaxy colour -- magnitude relation, calculated as defined in Sect.~\ref{sec:edge_measurement}. We can see that the bright edge evolves significantly with redshift. 
Galaxies in the blue cloud (i.e. those with $U-V \lesssim 1.6$), display a shift of the bright edge that is only weakly dependent on galaxy rest-frame colour, with a reduction in brightness of  $1.4$ to $1.7$ mag in the $V$ band. The edge evolves less for the red sequence, shifting faintwards by $0.7$ mag over the redshift range shown.  

The rest-frame $U-V$ colour correlates strongly with the SFR of a galaxy. 
However, the bright edge for blue galaxies appears to evolve as a coherent block. 
Hence, this seems to indicate that the shift of the bright edge of the colour -- magnitude relation is less sensitive to the instantaneous star formation rate. 
Of course galaxies with $U-V \gtrsim 1.6$ are already in a quiescent phase and hence display less of a change in colour than star-forming galaxies that are making the transition to the red sequence. 

In measuring the bright edge location we are dealing with the bright part of the galaxy luminosity function, so we do not have to worry about completeness effects in the VIPERS sample, for any redshift bin we consider. 
To show that this is indeed the case, i.e. that the bright edge is not driven by the luminosity cut at faint magnitudes, in Fig.~\ref{fig:cm_edge_evolution} we show, using thick grey lines, the location of the most populated bin in the magnitude distribution and we notice that it does not coincide with the faintest bin, ensuring that the depth of the survey has no effect on the definition of the bright-edge in the different redshift bins.

A concern could arise due to the limited volume that is sampled by VIPERS. Since bright objects are intrinsically rare, one might worry they could be missed at low redshifts, due to the lower volume sampled, compared with that probed at higher redshifts, for a fixed solid angle. In Appendix~B we assess the impact the volume probed has on the number of bright galaxies by subsampling the GALFORM output to contain the same number of objects as  VIPERS. We find that the bright edge is still in the same position, within the errors  (Fig.~\ref{fig:galform_zbins}).

\section{Evolution of synthetic SFH compared to observations}
\label{sec:results}

Making use of the modelling developed in Section~\ref{sec:cm_evolution}, we can synthetically evolve the observed VIPERS galaxies to different cosmic times, and hence see where they would appear in the colour -- magnitude plane in different redshift bins. We can then compare this synthetically evolved population with the observed population at the same  redshift and repeat this process for different redshift bins. Comparing the level of agreement between the bright edges of the colour -- magnitude relations for the synthetic and observed galaxies gives us some insight into simple SFH quenching scenarios. In the next sections, we explore the different SFHs in order from top to bottom from Fig.~\ref{fig:Sketch_SFH_99_0}. We consider first a scenario that does not involve any quenching (Section~\ref{sec:no_quenching}) and then two scenarios which impose quenching differently  (Section~\ref{sec:quenching_scenarios}): in the first the  quenching event occurs at the epoch of observation (Section~\ref{sec:quench_0}) while in the second the quenching takes place at a time that is randomly selected from selected time intervals after observation (Section~\ref{sec:quench_delayed}).

Note that our approach does not attempt to model the whole of the observed population at each  end-redshift. We are simply tracking the forward evolution of the VIPERS galaxies observed at $1.0<z<1.1$, assuming that the evolution is described by the best-fitting SFH to the observed photometry, with various quenching scenarios applied that truncate the star formation (see next sections). We assume that these galaxies preserve their identity from the observation redshift to the end redshift i.e. they do not merge with another galaxy. Also, we consider all of the evolved population of galaxies without requiring that the galaxies meet the VIPERS selection at the end-redshift.  
We remind the reader that our aim is not to reproduce the whole of the observed VIPERS colour-magnitude relation at redshifts below $z \sim 1$, but rather to focus on the evolution of the bright edge and how it compares to the observed one. Comparing the evolved and observed galaxy populations in the colour-magnitude plane is an interesting test which requires more sophisticated modeling. Moreover, such a test would be model-dependent and we feel that this goes beyond the scope of this paper. 

When comparing the synthetic and observed bright edges in the colour-magnitude plane, we need to bear in mind the following subtle point. The sample we are evolving using the best-fitting SFH, is defined at high redshift. VIPERS is a magnitude limited survey ($i_{\rm{AB}}<22.5$). Hence, galaxies fainter than this limit in the high redshift bin are not considered from the outset in the  analysis described in this section. This means that for substantial evolution times after the epoch corresponding to the high-redshift bin, the blue cloud will inevitably become depleted as  star-forming galaxies that would naturally enter the observed sample at lower redshifts are not considered in the modelling. This is because their progenitors in the high redshift bin were too faint to be included in VIPERS. For this reason, if we had found an excess of blue star-forming galaxies in the lower redshift bins this would indicate the need for a faster and more efficient mechanism to  suppress star-formation. However, the absence of blue galaxies when evolving a high-redshift sample, is expected and does not provide any new information about the nature of the quenching nor should it be interpreted as a mismatch between the observations and the modelling.

\subsection{The no-quenching scenario}
\label{sec:no_quenching}

The synthetic evolution of VIPERS galaxies observed at $1.0<z<1.1$ to lower redshifts is shown without any quenching in Fig.~\ref{fig:EVOLUTION_noquench}. The evolution is modelled using the best-fitting SFH to the observed photometry (corresponding to the schematic SFH in the top panel of Fig.~\ref{fig:Sketch_SFH_99_0}). 
The colour -- magnitude relation for this evolved set of galaxies is shown by the blue points in each panel of Fig.~\ref{fig:EVOLUTION_noquench}, along with the associated bright edge (solid blue line). 
Each panel in Fig.~\ref{fig:EVOLUTION_noquench} shows the evolved colour -- magnitude relation at a different time interval after the observation redshift ( left: $t_1-t_{\rm{obs}}=0.8$ Gyr, middle: $t_2-t_{\rm{obs}}=1.3$ Gyr,  and right:  $t_3-t_{\rm{obs}}=2.5$ Gyr). The observed VIPERS galaxies at each redshift plotted in Fig.~\ref{fig:EVOLUTION_noquench} are shown by the red points and their associated bright edge by the solid red line. Except for the reddest colour bins, there is a mismatch in the observed and synthetically-evolved bright edges, with the sense of the discrepancy depending on the colour. In the ``green valley'' ( $1.2 < U-V < 1.5$), the synthetic bright edge is fainter in the $V$-band than the observed one (by $\approx 0.5-0.75$ mag). This situation is reversed for blue galaxies ($U-V < 1.2$) for which the synthetic bright edge is $\approx 0.5$ mag brighter than the observed one.  

The excess of bright galaxies predicted with the no-quenching SFHs seems to point towards the need for a scenario with  widespread quenching of star formation activity for bright galaxies.
However, no conclusion can be reached about the SFH of fainter galaxies, i.e. those galaxies fainter than the peak of the distribution of luminosities (see Fig.~\ref{fig:histo_bright_edge}), as they do not feature in defining the location of the bright edge. 

Another important property of the quenching  suggested by Fig.~\ref{fig:EVOLUTION_noquench} is that it must take place over a wide range of colours, and hence affect observed galaxies with very different levels of star formation activity. Our stellar population modelling suggests that these galaxies also have a range of stellar masses, implying that stellar mass is not the only factor that governs the SFH of galaxies (see \ref{sec:masses}).

\subsection{Exploration of quenching scenarios}
\label{sec:quenching_scenarios}

Our aim is not to build a complete and physically motivated star formation quenching model, but rather to provide some indicative constraints on the quenching, as derived from the observed evolution of the galaxy properties in our sample. We therefore explore only a small number of simplistic quenching models (corresponding to the SFHs sketched in the lower two panels of Fig.~\ref{fig:Sketch_SFH_99_0}), to help us gain some insight into the quenching time-scales. We stress that the modified SFHs we have built to include quenching all assume that the quenching is complete, i.e. that no residual SFR remains after the quenching, and that the transition takes place almost instantaneously (effectively over a period of less than 100 Myr).

\subsubsection{Quenching the SFH at the epoch of observation}
\label{sec:quench_0}
The first and simplest quenching model we consider is one in which the best-fitting SFH to each galaxy is quenched immediately after the epoch of observation (i.e. corresponding to the schematic in the middle panel of Fig.~\ref{fig:Sketch_SFH_99_0}), thereby truncating the SFH and setting the subsequent SFR to zero. As we are using the high-redshift bin $1.0 < z < 1.1$ as our starting point, the quenching takes place at the redshift of observation for each galaxy. Even if the quenching redshift is very similar over the sample, this does not mean that the galaxies are at the same stage in their evolution, because their SFHs are described by different parameters. This can be easily seen from Fig.~\ref{fig:SFHs}. In fact, if we draw a vertical line at the same redshift, this will intersect every SFH at different stages in the evolution for different galaxies (i.e. $t$ will be similar for every galaxy but $t-t_{\rm{start}}$ will be different and for different SFH (different $\tau$) the evolutionary stage can be very different: the galaxy can be still star-forming or already in its passive phase).

Fig.~\ref{fig:EVOLUTION_quench0} shows, using the same format as Fig.~\ref{fig:EVOLUTION_noquench}, the evolution of the synthetic colour-magnitude relation for this SFH-quenching scenario.
We can see clearly from Fig.~\ref{fig:EVOLUTION_quench0} that instantaneous quenching is too extreme, resulting in the bright edge for synthetic galaxies being fainter than the observed one.  This is particularly noticeable for bluer galaxies.  
Again, as with the no quenching case, the exception is the reddest galaxies, for which the synthetic and observed bright edges coincide. These objects already have a SFR that is almost zero at the epoch of  observation, and therefore their SFH, and consequently their photometric properties, are not significantly affected by any quenching we might introduce.

\subsubsection{Delayed quenching}
\label{sec:quench_delayed}

To mitigate the excessive evolution of the bright edge found on quenching the SFH of all galaxies at the epoch of observation, we have explored delayed quenching scenarios, in which the quenching takes place at a time that is drawn uniformly between $t_{\rm{obs}}$ and $t_{\rm{obs}} + \Delta t_{\rm{quench}}$ (a schematic representation of this scenario is shown in the bottom panel of Fig.~\ref{fig:Sketch_SFH_99_0}). Another possibility would be to associate a quenching time delay with a specific galaxy property, but that would require us  to know which property might regulate the quenching. 

Given that the redshift interval covered by our sample corresponds to a cosmic time interval of $2.5$ Gyr, we have explored three values for the maximum delay time: $1$, $1.5$, and $2.5$ Gyr, as shown in the top, middle and bottom rows of Fig.~\ref{fig:EVOLUTION_quench123}. While the scenario with a maximum delay time of $1$ Gyr produces a bright edge evolution which is quite similar to that found with  instantaneous quenching, the two scenarios with maximum delay times of 1.5 or $2.5$ Gyr produce a bright edge which is in better agreement with the observed one. 

In each delayed quenching scenario explored we are testing the evolution at three end-redshifts which correspond to the cosmic times $t_1$, $t_2$ and $t_3$ in the bottom panel of Fig.~\ref{fig:Sketch_SFH_99_0}. For each of these times the fraction of quenched galaxies ($Q_{\rm{frac}}$ in Fig.~\ref{fig:EVOLUTION_quench123}) changes according to the scenario explored, with the exception of $t_3$ where we always have $100\%$ of the galaxy population quenched. 
We stress that $Q_{\rm{frac}}$ is an indirect consequence of the simple empirical quenching model that is useful to quote as it demonstrates one of the limitations of the model. As galaxies quench following a uniform distribution of times limited by $\Delta t_{\rm{quench}}$, $Q_{\rm{frac}}$ is simply defined by $Q_{\rm{frac}} = \Delta t / \Delta t_{\rm{quench}}$ which gives the probability that a galaxy has already experienced the quenching. We note that a $Q_{\rm{frac}}$ value of 100\% at a given redshift implies that there are no star-forming galaxies at lower redshifts, which disagrees with observations. 
For the quenching event spread over $\Delta t_{\rm{quench}}=1$ Gyr (top panel) we have $80\%$ of galaxies quenched at $t_1$ and $100\%$ at the other times. For $\Delta t_{\rm{quench}}=1.5$ Gyr, instead, we have $53\%$ of quenched galaxies at $t_1$ and $87\%$ at $t_2$. Finally, for $\Delta t_{\rm{quench}}=2.5$ Gyr, we have $32\%$ quenched at $t_1$ and $52\%$ at $t_2$.
If we focus exclusively on the bright edge location, and we define a good model as one that gives a good match between the observed and synthetic bright edges, then a scenario with $t_{\rm{quench}}=2.5$ Gyr is the best among the three explored.
This is particularly true if we focus our analysis only on $t_3$ (the right panels in Fig.~\ref{fig:EVOLUTION_quench123}), so that for each scenario (each row in Fig.~\ref{fig:EVOLUTION_quench123}) we are comparing a $100\%$ quenched sample. This suggests $2.5/2 =1.25$ Gyr as an estimate of the average quenching time-scale for our sample of galaxies.

\section{Contrasting the colour-magnitude evolution with galaxy formation models}
\label{sec:galform}

\begin{figure}
    \centering
    \includegraphics[width=0.70\textwidth]{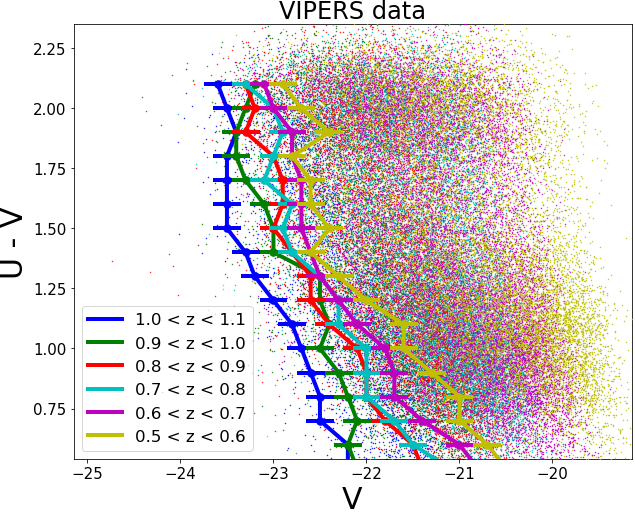}
    \vspace{1ex}
    \includegraphics[width=0.70\textwidth]{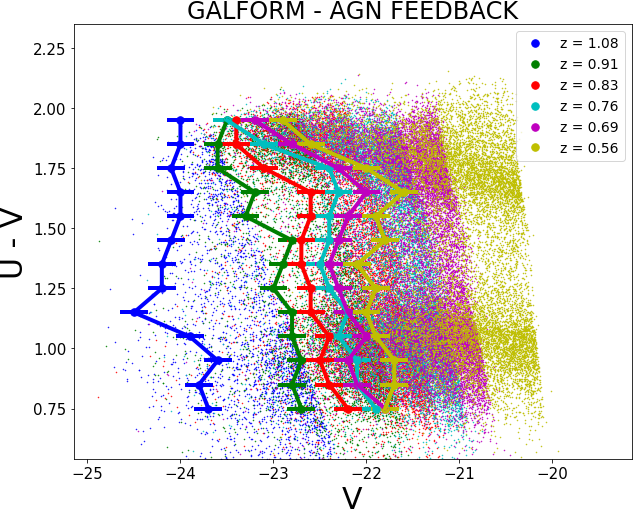}
    \vspace{1ex}
    \includegraphics[width=0.70\textwidth]{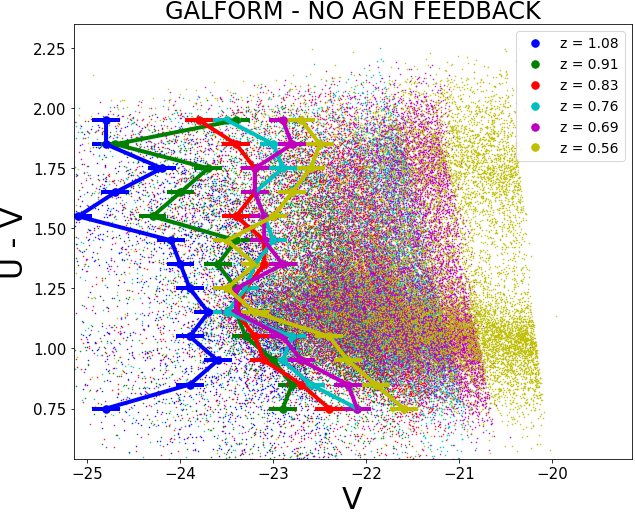} 
    \caption{\textit{Top panel:} VIPERS colour-magnitude diagram as in Fig.~\ref{fig:cm_edge_evolution} to be compared with the {\tt GALFORM} snapshots in lower panels. \textit{Middle panel:} GP14 {\tt GALFORM} colour-magnitude diagram (the default includes AGN feedback). A cut mimicking a $i_{\rm{AB}}\leq22.5$ cut (like VIPERS) has been applied to the model galaxies. \textit{Bottom panel:} same as middle panel except that AGN feedback has been turned off. The horizontal error bars show an indicative error of 0.15 mag}. 
    \label{fig:galform}
\end{figure}
The stellar population synthesis models and simple delayed exponential SFH -- quenching scenarios discussed in the preceding sections serve the purpose of developing some insight into galaxy 
evolution from the behaviour of the bright edge in the colour -- magnitude relation. Several studies of quenching have been performed using both semi-analytical (e.g.  \cite{hirschmann14,Henriques:2017})  and hydrodynamical  (e.g. \cite{Bluck:2016,Wright:2019}) models of galaxy formation.  
In this section we compare the VIPERS observations against the predictions of the semi-analytical galaxy formation model {\tt GALFORM} \cite{cole00,lacey16}. 

{\tt GALFORM} models a wide range of physical processes that govern the fate of the baryonic component in the universe, in the context of the hierarchical growth of the dark matter (for reviews see \cite{baugh06} and \cite{Benson:2010}). Of particular relevance to the quenching of star formation in galaxies is the radiative cooling of gas from hot halos. The cooling flow can be staunched by the luminosity of an active galactic nucleus \cite{bower06}. In simple terms this occurs when the cooling time of the hot halo gas exceeds the free-fall time and the cooling luminosity is balanced by the energy released by material falling onto the supermassive black hole at the centre of the galaxy. Star formation in a galaxy for which there is no cooling flow would be truncated once the existing cold star-forming gas is consumed and if it is not replenish by cold gas brought in by a merger. The aim here is to compare the colour -- magnitude relation predicted by a standard version of {\tt GALFORM} with the VIPERS observations. Exploiting the speed and modularity of semi-analytical models, we also examine a variant of the fiducial model in which we turn off AGN feedback to see the impact on the model predictions.

The {\tt GALFORM} model that we consider here is the version introduced by \cite{gonzalez14}, hereafter GP14. This model is implemented in a version of the Millennium N-body simulation that adopts a cosmology consistent with the 7 year WMAP results \cite{guo11}. The GP14 model assumes a standard solar neighbourhood stellar initial mass function (IMF) \cite{kroupa98} for quiescent and burst star formation, which is the main feature which distinguishes it from the dual IMF models of \cite{lacey16} and \cite{Baugh:2019}. The parameters of the GP14 model have been calibrated to reproduce a range of observations, including the $z=0$ galaxy luminosity function in the $b_{\rm J}$ and $K$ bands. The luminosity function predicted by {\tt GALFORM} has been compared with observations in the optical to $z\sim0.3$ by \cite{mcnaught14} for red and blue galaxies and for different environments, using the Galaxy And Mass Assembly (GAMA) survey. These authors found that the model predictions agreed reasonably well with the observations for bright galaxies, with the biggest discrepancy being the over prediction of faint-red galaxies, which can be traced to the treatment of gas cooling in satellite galaxies (see the discussion in \cite{Font:2008}) \footnote{Similar agreement with observed luminosity functions is found in other semi-analytical models (e.g. \cite{hirshmann16,Henriques:2017,Lagos:2019})}. 
The N-body simulation is a cube of volume $(500\, {\rm Mpc}/h)^3$ which we note in passing is substantially bigger than the effective volume of VIPERS, which is equivalent to a box of volume $(368\, {\rm Mpc}/h)^3$.

We consider the GP14 model at simulation outputs that are close to the centres of the VIPERS redshift bins. We apply the VIPERS apparent magnitude cut of $i_{\rm{AB}}\lesssim 22.5$. Using {\tt GALFORM} snapshots at precise redshifts rather than VIPERS redshift ranges results in a sharp well-located cut in the {\tt GALFORM} colour-magnitude plane at faint magnitudes. However, since the focus of the test is on the bright-edge, this does not introduce additional uncertainties. The majority of the model galaxies selected in this way are central galaxies in intermediate to massive halos, accounting for at least $75\%$ of the total at all redshifts. Satellites galaxies without an identifiable  dark matter subhalo represent a small fraction of the sample, reaching a peak of $8\%$ at redshift $0.56$ for colours redder than $U-V>1.5$, leaving the evolution of the blue cloud unaltered.

The primary quenching mechanism for central galaxies is AGN feedback (in particular the hot halo -- radio mode AGN feedback), which shuts down gas cooling in halos above a mass that is determined by the parameters describing the AGN feedback (see \cite{lacey16}). This motivates our choice to explore a variant model in which this quenching is turned off, by switching off AGN feedback (by setting $\epsilon_{\rm{heat}}=0$ and leaving the other parameters unchanged; see section 3.5.3 in \cite{lacey16}). Note that this is not a viable model as it produces too many bright galaxies. Nevertheless it is illustrative to see the impact on the colour -- magnitude relation of turning off this quenching mechanism. 

Note that AGN feedback is not the only star formation quenching mechanism in {\tt GALFORM}. Satellite galaxies have their star formation quenched by stripping of their hot gas halo after they are accreted into a more massive dark matter halo. This stops any further gas cooling onto the satellite, which leads to less star formation than may have been the case if the galaxy had remained a central galaxy. However, the effect of this process on the colour -- magnitude diagram is negligible as central galaxies dominate the {\tt GALFORM} sample for the VIPERS selection.
Also, the quenching due to AGN feedback in {\tt GALFORM} is not as dramatic as that in the simple model considered in earlier sections of this paper as {\tt GALFORM} galaxies do not stop their  star formation instantaneously but in a gradual way as the molecular gas reservoir is exhausted.
Bright galaxies in the {{\tt GALFORM}} model, which have had their quiescent star formation quenched by the suppression of gas cooling by AGN feedback, can still display episodes of star formation activity through bursts involving the cold gas brought in by merging galaxies.

The evolution of the colour -- magnitude relation in the observations and models is compared in Fig.~\ref{fig:galform}. The VIPERS observations in different redshift bins are collected in the upper panel to give a different view of the evolution of the bright edge from that shown in Fig.~\ref{fig:cm_edge_evolution}. The middle panel shows the predictions of the GP14 model and the lower panel the variant of GP14 with no AGN feedback. The rest frame magnitudes and colours used in the {\tt GALFORM} predictions have been chosen to match the transmission curves of the filters used in VIPERS and include attenuation by dust (see \cite{GP:2013} and \cite{lacey16} for a description of the implementation of attenuation by dust in {\tt GALFORM}).
To make the {\tt GALFORM} colour-magnitude plot more visually comparable to those from VIPERS we randomly subsample the model galaxies to account for the difference in volume between the {\tt GALFORM} simulation and VIPERS.
 
The top panel of Fig.~\ref{fig:galform} shows the steady evolution of the bright edge to fainter luminosities with increasing redshift in the VIPERS observations. This evolution is strongest in blue $U-V$ colour bins, with the bright edge becoming approximately 1.5 mag fainter from $z \sim 1$ to $z \sim 0.5$, and weakest for red colour bins, with the edge being around 0.75 mag fainter over the same redshift interval.

The middle panel of Fig.~\ref{fig:galform} shows the colour -- magnitude relation for the fiducial GP14 {\tt GALFORM} model. The first thing we notice on comparing with the VIPERS data is the overall shift in the locus of the model points to brighter magnitudes, reflecting the difference between the {\tt GALFORM} and VIPERS luminosity functions.  \cite{fritz14} measured the evolution of the luminosity function using VIPERS and find a sharp break at the bright end, across the redshift range considered here. Our comparison of the \cite{fritz14}, results with the {\tt GALFORM} predictions shows {\tt GALFORM} produces a weaker break, and hence more bright galaxies than estimated from VIPERS, with a dip in the model luminosity function relative to that from VIPERS around $L_*$. 

Focusing on the relative change in the bright edge in the {\tt GALFORM} predictions rather than its absolute position, Fig.~\ref{fig:galform} shows that the model bright edge also moves faintwards with decreasing redshift, preserving the ordering of the bright edge with redshift more clearly than the data. The shift in the position of the bright edge for the bluest colour bins is about $0.5$ mag larger than for VIPERS. 
The shift of the edge for the reddest bin is just over 1 mag, approaching twice the shift for VIPERS. 

The shift of the bright edge towards fainter magnitudes with time is an interesting feature which deserves some discussion here. The key point to bear in mind when interpreting this shift is that the galaxies which define the bright edge are not the same ones at every redshift. 
The bright edge galaxies have lower stellar masses as we move to lower redshifts. Of course the stellar mass of an individual galaxy increases throughout its lifetime. However, as the star-formation in these galaxies declines, they will become fainter, and will no longer be part of the 15 per cent of objects in a given colour bin which define the location of the bright-edge. Moreover, after sufficient time, they will become fainter than the limiting magnitude of the survey ($i_{\rm{AB}}<22.5$) and hence will not appear in the colour - magnitude diagram. The shift of the bright edge to fainter magnitudes with time, observed in both VIPERS and the {\tt GALFORM} model, reflects the effect of downsizing described in the introduction: lower mass galaxies are the ones most actively forming stars at low redshift, whereas at higher redshift it is the higher mass galaxies. 

Finally the lower panel of Fig.~\ref{fig:galform} shows the predictions of the variant of GP14 with AGN feedback switched off. The locus of model galaxies is shifted to brighter magnitudes, even compared to the fiducial GP14 model shown in the middle panel. This reflects the additional gas cooling in intermediate and massive halos compared to the fiducial model with AGN feedback switched on. This emphasises that this variant is not a viable model as it produces too many bright galaxies. The ordering of the bright edges with redshift is also lost, with the bright edges for several redshifts overlapping for the blue cloud.

Another notable feature of the {\tt GALFORM} predictions in  Fig.~\ref{fig:galform} is the strong bimodality in colour, as pointed out by \cite{bower06} and \cite{gonzalez09}. 
This is also apparent in Fig.~\ref{fig:EVOLUTION_noquench}, when we use smooth delayed exponential SFHs without quenching. 

In addition, we note that in the VIPERS data the bright edge is vertical for red colours, and becomes increasingly diagonal moving to bluer colours. The edge shifts as a coherent block back to lower luminosities with time. The fiducial {\tt GALFORM} model does not reproduce this behaviour. The bright edge is vertical for the blue sequence, and diagonal for the red sequence. In the VIPERS data, the bright edge traces the blue edge of a diagonal blue sequence, while in the GALFORM output, it traces the bright-edge of a horizontal blue sequence. 

Whilst there are qualitative differences between the colour-magnitude relations of VIPERS and the {\tt GALFORM} model (e.g. shift to brighter magnitudes, the shape of the bright edge and stronger colour bimodality in {\tt GALFORM} than in VIPERS, at approximately $U-V\sim 1$ and $U-V\sim 1.75$), the ordering of the bright edges with redshift and the breaking of this ordering in the variant model without AGN feedback indicate that AGN feedback plays an important role in quenching star formation activity in the models.

\section{Discussion}
\label{sec:discussion}

The very notion of quenching star formation in galaxies is somewhat dependent on the galaxy evolution paradigm adopted to interpret observations or build models, as discussed extensively by \cite{abramson16}. It is possible that the need for quenching is merely the result of incorrectly assuming that star formation is a simple process. However, as this simplifying assumption is often the first step in any attempt to model galaxy evolution, it is worthwhile to try and characterise any quenching that may take place. 
 
The need for quenching remains controversial. \cite{CIESLA_16} advocated for the quenching of star formation on the basis of SED analysis of the Herschel Reference Survey galaxy sample \cite{boselli10}, applying delayed exponential SFHs, similar to those used here, and adding a sharp quenching to the SFR\footnote{The only difference with our model is that the quenching invoked by \cite{CIESLA_16} is not complete, but reduces the SFR to $35$\% of its prior value.}. On  the other hand \cite{pozzetti10} claimed that the colour evolution in the zCOSMOS sample could only be reproduced by including extended SFHs or secondary bursts of star formation in the SED modeling, rather than by a reduction or suppression in the level of star formation activity. One reason for this disagreement might be the SFH modelling  adopted by \cite{pozzetti10}, which differs in two key ways from that adopted here, namely that all galaxies follow exponentially declining SFHs\footnote{ ${\rm SFR}\left(t,\tau,t_{\rm{start}}\right) = \left( 1/\tau \right) \exp{ \left[ -\left(t-t_{\rm{start}}\right) /\tau \right]}$.}, and that no evolution of the dust content takes place inside galaxies as their star formation evolves. Some studies \cite[see for example the discussions in][]{citro16,tomczak16,abramson16} claim that the use of purely exponentially decreasing SFHs is not ideal for studying the evolution of galaxy properties as this does not allow for an initial increase of the SFR. The exponentially declining SFH results in reddening of galaxy colours by construction, without the flexibility to describe the population of star-forming galaxies which are still moving towards bluer colours and brighter magnitudes.

 Much recent work on quenching has focused on a global description of the galaxy population, often using the stellar mass function, as in \cite{bundy06} and \cite{peng10}, or measuring the transition of galaxies across the green valley, as in \cite{schawinski14} and \cite{lian16} (see also \cite{Wright:2019} for a measurement of a quenching time based on the green valley and other techniques relying on the time at which galaxies leave the main sequence of star formation, for a hydrodynamical simulation of galaxy formation). 
 Here instead we study how different SFHs both with and without a truncation of the star-formation activity can affect the bright edge of the colour-magnitude plane, over a substantial lookback time, corresponding to the redshift range $0.5\lesssim z \lesssim 1.1$.

We use two approaches to interpret the evolution of the bright edge in the VIPERS colour -- magnitude relation. The first is a simple empirical model in which the evolution of VIPERS galaxies is predicted using a smooth SFH that is quenched at or after the epoch of observation. The SFH is a delayed exponential that is the best-fitting description of the galaxy's photometric properties (rest frame $U-V$ colour and rest frame $V$-band absolute magnitude). The second approach is a physically motivated semi-analytical model of galaxy formation, which predicts the evolution of the colour -- magnitude relation directly. 
 
The SFHs predicted by physical galaxy formation model appear to be much more complicated than the smooth delayed exponentials adopted in the empirical model (see the examples from {\tt GALFORM} plotted in \cite{baugh06}).  \cite{simha14} compared a range of parametric SFHs to those predicted in a hydrodynamic simulation of galaxy formation. Their ``lin-exp'' model, which is equivalent to the delayed-exponential used here, performed much better overall, experiencing problems mainly for the very bluest and reddest galaxies. Without a truncation of the SFH, lin-exp models result in higher values of SFR at early times and lower values of SFR at late times (as any truncated SFH would be interpreted as a smooth SFH with a very small $\tau$\footnote{The limitation of having SFHs defined by only one parameter, $\tau$, is that early and late times in the life of a galaxy are related. Introducing a quenching event overcomes this limitation.}). The use of contrasting approaches to model the evolution of the colour -- magnitude relation lends robustness to any consistent conclusions reached about the importance and nature of quenching.
 
Our simple empirical analysis of the evolution of the bright edge of the colour -- magnitude relation suggests that quenching must be a widespread phenomenon, taking place over the full range of redshift  ($0.5<z<1.1$) and stellar mass (approximately $9.0 \lesssim \log(\mathcal{M}/M_\odot) \lesssim 11.0 $) probed by VIPERS. 
The range of stellar masses that is quenched is an interesting result. Following the identification by \cite{kauffmann03} of a threshold stellar mass ($\sim 3 \times 10^{10} M_{\odot}$) above which local galaxies are dominated by passive early-types, the concept of a transition mass above which galaxies are quenched has become popular \cite[see for example][]{bundy06,davidzon13}. However, our results imply instead that quenching takes place over a wider range of stellar masses than what proposed by \cite{kauffmann03}. (See \ref{sec:masses} for the justification of the range of stellar masses involved.) Other studies such  as \cite{schreiber15} study the evolution of the main sequence of star formation with redshift, and their results are consistent with quenching over a wide range of stellar masses.

The detection of the widespread quenching presented in Section~\ref{sec:no_quenching} is a robust result of this work, since the SFH modelling it is based upon reproduces quite well the main characteristics of star formation activity in the VIPERS galaxy sample, including the instantaneous measurement of SFRs at all redshifts and colours (see \ref{sec:sfh}).
The characterisation of the quenching time-scale that we discuss in Section~\ref{sec:quenching_scenarios} is instead based on a simple toy model, with the main purpose of demonstrating that a viable quenching history capable of describing the observations does indeed exist.

Finally, as a consequence of using a simplified SFH model, we do not attempt to identify a single galaxy property to replace stellar mass as the clearest driver of quenching and overall galaxy evolution. We note, however, that \cite{haines17} discuss at length the possibility that this property could be the mean stellar mass density, i.e. the amount of galaxy stellar mass located within the galaxy central kiloparsec. Of course our reliance on the evolution of the bright edge of the colour-magnitude relation to characterize quenching prevents us from drawing reliable conclusions about the possible quenching of galaxies that, at any epoch, are significantly fainter than the bright edge (i.e. fainter than the magnitude at which the number of galaxies in a given colour bin peaks). Still, the very simplistic assumption we make that only the galaxies whose evolution we are able to constrain (i.e. those that are close to the bright edge in terms of magnitudes) are actually undergoing quenching over a limited time-span after observation matches the results of \cite{lian16}, who estimate that only approximately one quarter of the galaxies in their sample start the quenching transition every gigayear. \cite{lian16} quench  their SFHs in a somewhat less dramatic way than we do. These authors use what they call a ``two-phase exponentially declining SFH", with one exponential describing the secular star-forming stage and the other describing a rapid quenching stage (see Fig.~1 in \cite{lian16}). Although the quenching is introduced as an exponential decline, their approach is not very different from ours as they estimate a quenching e-folding time of $500\, \rm{Myr}$ while our truncation is limited in time by the resolution of our models which is $100\, \rm{Myr}$. With these assumptions, \cite{lian16} study the drop in the number density of the  NUV$-u$ vs $u-i$ colour-colour diagram, finding a time to migrate from the star-forming to the passive population of $1.5 \,\rm{Gyr}$ (Fig.~2 of \cite{lian16} shows how these two populations are defined in the NUV$-u$ vs $u-i$ colour-colour diagram).

An indication of a physical processes that could explain the quenching of galaxy star formation rates since the peak epoch of global star formation was offered by  
the comparison of the predictions of the {\tt GALFORM} semi-analytical model of galaxy formation with the VIPERS colour -- magnitude relation. The qualitatively similar evolution of the bright edges in {\tt GALFORM} and the observations, and the break down of this evolution when AGN heating of cooling gas is turned-off by hand, reveals that AGN feedback heating is quenching the SFR by turning off the fuel supply for star formation. The bright edge evolves too much in {\tt GALFORM}, which might point to the need to revise the treatment of the reincorporation of gas heated by supernovae, as argued by \cite{Mitchell:2016} in their study of the evolution of the stellar mass -- halo mass relation.

In conclusion, our work points towards a SFH scenario in which quenching of star formation gives a better match to the evolution of bright galaxies in the colour -- magnitude plan that is not reproduced in models in which star formation is able to proceed unfettered.

\section*{Acknowledgements}
GM is supported by a PhD Studentship with the Durham Centre for Doctoral Training in Data Intensive Science, funded by the UK Science and Technology Facilities Council (STFC, ST/P006744/1) and Durham University. 
CMB and PN acknowledge support from the STFC through ST/P000541/1. PN
acknowledges the support of the Royal Society through the award of a University Research Fellowship. This work used the DiRAC@Durham facility managed by the Institute for Computational Cosmology on behalf of the STFC DiRAC HPC Facility (www.dirac.ac.uk). The equipment was funded by BEIS capital funding via STFC capital grants ST/K00042X/1, ST/P002293/1, ST/R002371/1 and ST/S002502/1, Durham University and STFC operations grant ST/R000832/1. DiRAC is part of the National e-Infrastructure.

Special thanks go to Micol Bolzonella for organising weekly teleconferences crucial for the development of this work and to Bianca Garilli for  constructive scientific discussions.

MS has been supported by the European Union's  Horizon 2020 research and innovation programme under the Maria Sk{\l}odowska-Curie grant agreement No 754510 and National Science Centre (grant UMO-2016/23/N/ST9/02963).

We acknowledge the crucial contribution of the ESO staff for the management of service observations. In particular, we are deeply grateful to M. Hilker for his constant help and support of this program. Italian participation in VIPERS was funded by INAF through the PRIN 2008, 2010, and 2014 programs. 

\bibliography{manzoni.bib}

\begin{appendix}

\section{SFR from SED and [OII] emission}
\label{sec:sfh}
\begin{figure}
\centering
\includegraphics[width=0.8\textwidth]{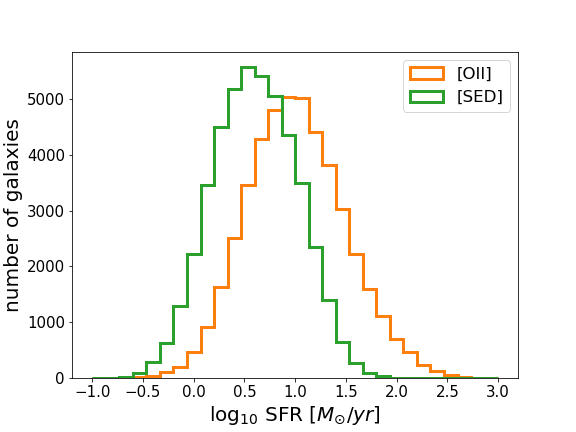}
\caption{Distribution of the SFR estimates from the SED fitting (green line) compared to those derived from the [OII]$\lambda$3727 line (orange line) according to the \protect\cite{MOUSTAKAS_14} prescription, as in Eq.~\ref{eq:SFR_OII}. In this sample we have excluded galaxies with signal-to-noise lower than $1.1$.}
\label{fig:SFR_histos}
\end{figure}
\begin{figure}
   \centering
    \includegraphics[width=0.8\textwidth]{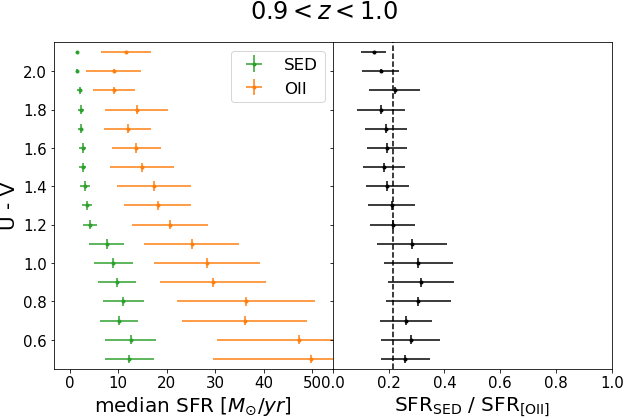}
    \par\medskip
    \includegraphics[width=0.8\textwidth]{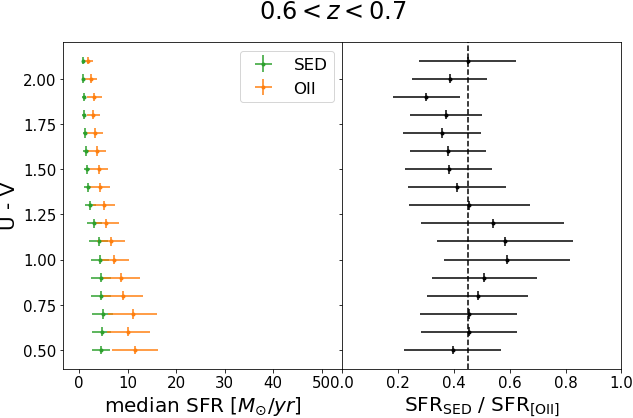}
     \caption{The distribution of SFR as a function of rest-frame colour for galaxies in the redshift ranges $0.9<z<1.0$ (top panels) and $0.6<z<0.7$ (bottom panels), as estimated from the OII line luminsity and inferred from SED fitting, as labelled. On the right hand side, the ratio between the two estimates of SFR is plotted on the $x$-axis for bins of colour plotted on the $y$-axis. The points show the median values and error-bars are obtained from the median absolute deviation (MAD) of the relative distribution.}
         \label{fig:SFR_vs_color}
   \end{figure}


Here we perform a ``sanity check'' to test if the SFR deduced from the SED fitting to the  $U-V$ colour tracks the SFR inferred independently from an emission line. This test has the limitation of testing the instantaneous SFR and does not probe the full SFH of galaxies. However, the importance of this test is the fact that we are estimating the same quantity from completely independent properties, i.e. the SED fitting is based on the spectrum sampled using broad filters  which is completely independent from the intensity of an emission line.

We first compare the SFR inferred for the star-forming galaxies in the observed sample, estimated using the [OII]$\lambda$3727 emission line luminosity, with that estimated from the properties of the best-fitting SED template. The prescription to obtain SFR estimates from the [OII] luminosity comes from \cite{MOUSTAKAS_14}. 
Specifically, from Table~2 of \cite{MOUSTAKAS_14} we made a linear fit to the coefficients $M_B$ and $P_{50}$ (neglecting the two faintest magnitudes which are outliers), resulting in the following relation between [OII] luminosity and SFR:

\begin{equation}
\label{eq:SFR_OII}
    \log_{10} \left(\frac{\rm{SFR}}{M_{\odot}/\rm{yr}}\right) = -2.893 -0.169 \times M_B + \log_{10} \left(\frac{L[OII]}{10^{41}\,\rm{erg}/\rm{s}}\right)
\end{equation}
where $M_B$ is the rest-frame absolute magnitude in the $B$-band and $L[OII]$ is the luminosity of the OII line. 

In Fig.~\ref{fig:SFR_histos}, we show the distribution of SFRs derived from the [OII] line luminosity (orange line) compared to the distribution of SFRs derived from the SED fitting (green line). There are several physical reasons why we do not expect these estimates to be exactly the same. One is dust extinction. The [OII] line may suffer from additional dust extinction compared to that experienced by the stellar continuum that is modelled in the SED fitting. Another reason may come from the different star formation time-scales sampled in the two approaches. \cite{KENNICUTT_13} suggests that the [OII] emission line samples star formation time-scales $\lesssim 20$ Myr, while in our SED fitting, star formation is sampled for interval times of the order of $100-200$ Myr. For this reason, the [OII] line is more likely to sample galaxies which are experiencing a starburst rather than being consistent with a smooth SFH as is the case with the templates used in the SED fitting. 

A more accurate test is to check if the SFR inferred from the SED fitting displays the same qualitative behaviour as the SFR estimated from the [OII] emission line. With this aim, we split the sample by $U - V$ colour and redshift. In Fig.~\ref{fig:SFR_vs_color}, the median SFR is plotted for the same colour bins used in defining the bright edge of the colour -- magnitude relation. The top and bottom panels in the figure show the same analysis in two redshift bins. The right-hand-side panels show the median ratio between the two estimates. The error bars indicate the Median Absolute Deviation (MAD) dispersion around the median values. We use the MAD to estimate the dispersion in the SFR values and ratios distribution because of its robustness against the presence of outliers in the distribution and insensitivity to the particular choice of parameters. What ensures the quality of the SFR estimates from the SED fitting is the fact that, although different from the [OII] SFR estimates, they follow the same trend both in colour (different points of the y-axis) and  redshift (top and bottom panels). This can be seen from the fact that the distribution of the ratios is constant within the errors (right-hand-side panels). 
The value of the mean ratio changes somewhat between the two redshifts considered and this could be due, for example, to an underlying  change in metallicity of the gas involved in the star formation. However, this does not affect our results as what is important is that within every redshift bin the distribution of the ratios is constant within the errors as can be seen from the right-hand  panels. 
In particular the relation between redshift and SFR shown in Fig.~\ref{fig:SFR_vs_color} assures us that the overall decrease in star formation rate density as a function of cosmic time that took place over the redshift interval $0<z<2$ is well reproduced.

Considering that our sample includes galaxies spanning more than two decades in stellar mass and two decades in star formation activity strength (with a strong dependence on redshift), we argue that our stellar population synthesis effort provides consistent estimates of the ``visible'' SFR within the VIPERS sample (i.e. the star formation measurable using optical emission lines or broadband colours).

Although this test does not provide evidence about the future evolution of the SFR, the consistency in reproducing qualitatively the SFR-colour and SFR-redshift fundamental relations give us an indication that the SED fitting is based on stellar population synthesis templates which provides a realistic starting point in our evolutionary exercise.

\section{Robustness of the bright edge to the number of galaxies}
\label{app:galform}
   \begin{figure*}
   \centering
    \includegraphics[width=\textwidth]{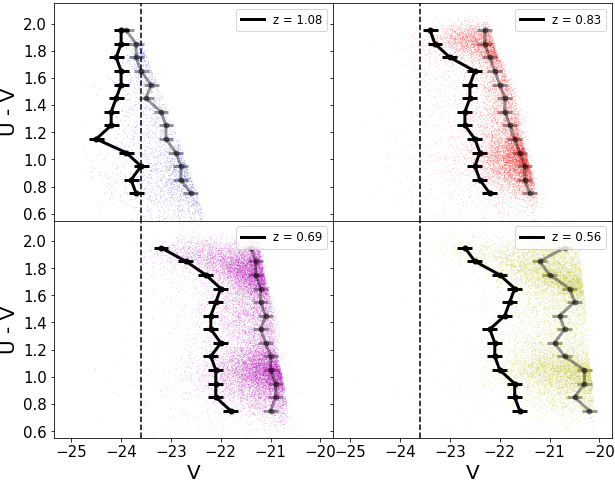}
    \includegraphics[width=\textwidth]{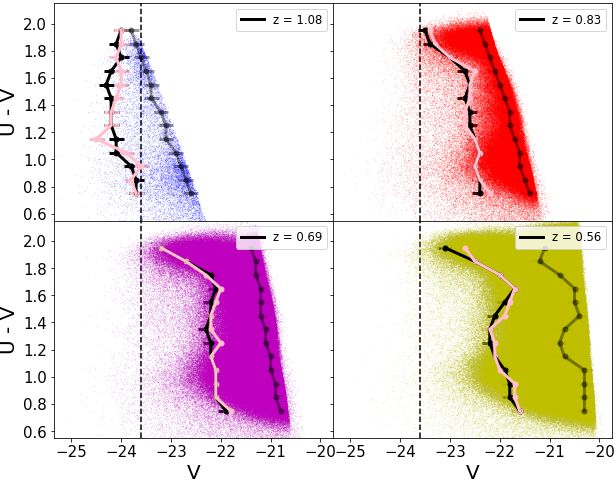}
     \caption{Colour-magnitude relation for GP14 {\tt GALFORM} semi-analytic model. The top panel is the randomly sampled version of the bottom panel in order to have the same number of points as in the corresponding VIPERS redshift bins. It can be seen that the definition of the bright edge is consistent in the two cases. For simplicity only four snapshots have been plotted. As in Fig.~\ref{fig:cm_edge_evolution}, the grey line represents the most magnitude populated bin for every colour bin. In the bottom panel, the edge relative to the sub-sampled sample has been reported for comparison with pink lines.}
\label{fig:galform_zbins}
   \end{figure*}
One potential worry is that the location of the bright edge as defined in Sect.~\ref{sec:edge_measurement} is affected by one of the survey properties, such as the number density of galaxies in the colour -- magnitude diagram. In Section~\ref{sec:galform} we have used a sub-sampled version of the original GP14 {\tt GALFORM} output in order to match the number of VIPERS galaxies in every redshift bin. We now want to compare this sub-sampled set of model galaxies to the original sample to see if this affects the location of the bright edge. In Fig.~\ref{fig:galform_zbins} we plot the colour -- magnitude relation and the associated bright-edge for 4 representative snapshots. In the top panel, we show the sub-sampled version, which is the one presented in the middle panel of Fig.~\ref{fig:galform}. In the bottom panel instead we use all the data available from the GP14 simulation. Albeit with less scatter, the location of the bright edge is consistent to the one in the top panel within the errors.
As in Fig.~\ref{fig:cm_edge_evolution} we also draw the location of the most populated magnitude bin, to show that is not coincident with the faint magnitude cut, hence the survey depth does not affect the location of the bright edge.\\

We have done an additional check to prove that the change in cosmic volume sampled by VIPERS at different redshifts is not the factor responsible for the progressive disappearance of the rare bright galaxies from the sample, and hence to prove that the bright edge evolution is not a volume sampling effect. 
The volume sampled in the highest redshift bin is in fact approximately ten times bigger than the lowest redshift bin.

To test for this possibility we have focused on the sample in the highest redshift bin, and computed how many of the galaxies in this sample would still be observed at a lower redshift, under the assumption of no evolution in the galaxy properties but just taking into account the ratio of the volumes sampled in the two different redshift bins. 
The result of this computation shows that the high redshift edge would still be visible down to $z\simeq 0.5$, if only volume effects were to dominate the VIPERS sample composition. 
This test therefore demonstrates that the observed bright edge evolution with redshift is real, and must be due to the evolving properties of the galaxies in the sample.

\section{Comparison with zCOSMOS}
\begin{figure}
   \centering
    \includegraphics[width=\textwidth]{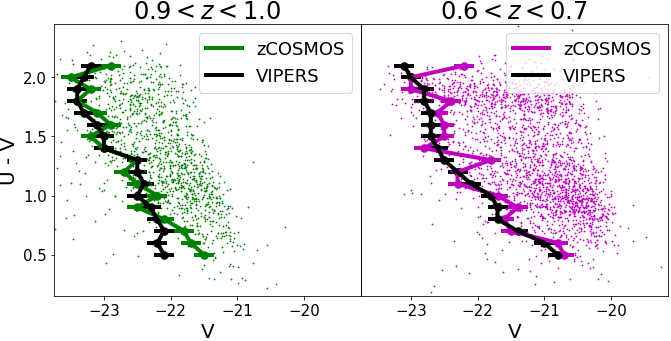}
     \caption{Comparison of the colour-magnitude bright edges between zCOSMOS and VIPERS data. The coloured points and edges come from the zCOSMOS catalogue while the black edges are relative to the VIPERS data in the same redshift bins (as plotted in Fig.~\ref{fig:cm_edge_evolution}). Only two redshift bins are shown here, but the matching between the two samples is of similar quality for all bins.}
         \label{fig:ZCOSMOS_67}
   \end{figure}

One further test of our analysis consists of verifying whether the same evolution of the colour-magnitude bright edge is observed in other redshift surveys.
We have taken the available data from the zCOSMOS bright sample \cite{lilly09}, divided the sample into the same redshift bins used for the VIPERS sample, and computed the zCOSMOS bright edge location using the same procedure described in Sect.~\ref{sec:edge_measurement}. Note that zCOSMOS has the same depth of VIPERS, i.e. it has been cut at magnitude 22.5 in the observed $i$-band. The difference with VIPERS is that zCOSMOS is a purely magnitude limited sample whereas VIPERS has a colour-colour pre-selection to isolate galaxies with $z\gtrsim 0.5$ (see Fig.~3 of \cite{guzzo14} for details). As can be seen in Fig.~\ref{fig:ZCOSMOS_67} (where we plot only two redshift bins for simplicity), the VIPERS and zCOSMOS bright edge locations coincide almost perfectly, albeit the zCOSMOS one is somewhat noisier because of the smaller number of objects in that sample, since the area covered by zCOSMOS is significantly smaller than the one covered by VIPERS. Although only two redshift bins are shown, the agreement is of the same quality for all of the redshift bins used in our VIPERS study.

\section{Stellar masses in VIPERS}
\label{sec:masses}

\begin{figure}
\centering
\includegraphics[width=\textwidth]{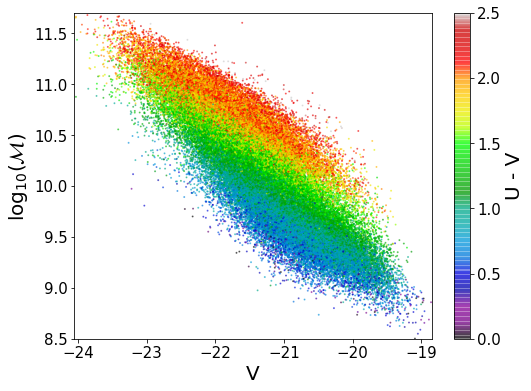}
\caption{Relation between the rest frame V magnitude and stellar masses, colour coded by the $U - V$ colour. All stellar masses are in units of M$_{\odot}$.} 
\label{fig:stellar_masses}
\end{figure}
   
\begin{figure*}
   \centering
    \includegraphics[width=\textwidth]{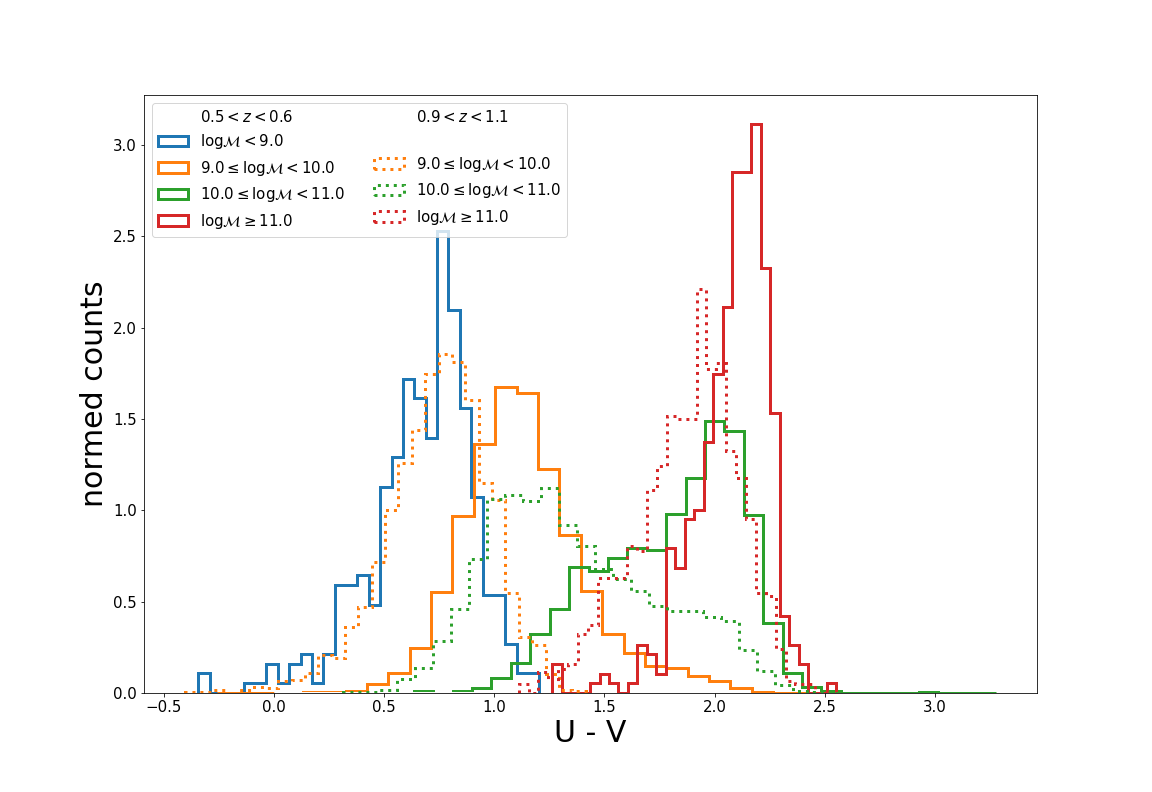}
     \caption{Histogram of the $U-V$ colour distribution split in 4 mass bins indicated by the colour of the lines, and 2 redshift bin indicated by the shape of the line. All stellar masses are in units of M$_{\odot}$.}
    \label{fig:stellar_masses_histo}
\end{figure*}

The aim of this section is to show how the bright edge of the colour - magnitude relation is made up of galaxies that span a wide range of stellar masses. Here stellar masses are obtained as in \cite{moutard16b}. The evolution of the bright edge points toward a scenario where quenching is acting on galaxies which are bright in the $V$-band but which have different stellar masses. To show this, we plot in Fig.~\ref{fig:stellar_masses} the relation between the rest frame $V$-band magnitude and stellar mass in the VIPERS sample. We use different coloured symbols in Fig.~\ref{fig:stellar_masses} to differentiate between galaxies with different $U - V$ optical colours. 
We observe that the brightest red galaxy (e.g $U-V = 2$) and the brightest blue galaxy (e.g. $U-V = 0.5$) not only have different V-band absolute magnitudes ($V\sim -23$ for the bright red galaxy compared with $V\sim-21.5$ for the bright blue galaxy), but also different stellar masses.
We can also look at the distribution of $U-V$ optical colours in Fig.~\ref{fig:stellar_masses_histo}. In this plot we split the analysis into different stellar mass (different line colours) and redshift bins (different line style). We notice that higher values of stellar mass dominate at redder $U-V$ colours while lower stellar masses dominate at bluer colours. This holds true both in our low redshift bin ($0.5<z<0.6$, solid line histogram) and in our high redshift bin ($0.9<z<1.1$, dashed line). 
Moving from high to low redshift we see an evolution in the $U-V$ colour, with galaxies  becoming systematically redder but the order of the stellar mass bins is preserved in colour, with lower stellar masses always being bluer than higher stellar masses. We can state that, in our highest redshift bin, galaxies are more massive and bluer. 
In fact at $0.9<z<1.1$ we do not find any galaxy in our sample with $\log \mathcal{M} < 9.0$. 
Combining the information from Figs.~\ref{fig:stellar_masses} and~\ref{fig:stellar_masses_histo}, we can conclude that, over the  redshift range ($0.5<z<1.1$),  galaxies that are bright in the $V$-band, which are those that define the bright edge, have a wide range of stellar masses (independently of the colour evolution). 
Since  galaxies approaching the bright edge are those which stop becoming bluer and brighter, and start getting fainter and redder, we argue that the turn over is due to quenching and that this mechanism affects a variety of stellar masses. In Section~\ref{sec:galform}, we illustrate the impact of switching off AGN feedback 
on the colour-magnitude relation. This is a process that mainly affects the cooling onto central galaxies in massive halos, and so without AGN feedback there is an increase in the abundance of bright galaxies. However, we do not expect AGN feedback to have an impact on the galaxies in low mass haloes. Other processes which restrict star formation, such as feedback from SNe and the stripping of hot gas from galaxies that fall into clusters will also act to suppress star formation, particularly at lower masses. A study of how AGN and SNe feedback can change the stellar mass of a galaxy is presented in \cite{Mitchell:2016}.


\end{appendix}

\end{document}